\documentclass[12pt]{article}

\usepackage{graphicx,psfrag,epsf}
\usepackage{enumerate}
\usepackage{natbib}
\usepackage{mathtools}
\usepackage{booktabs}
\usepackage{color}
\usepackage[english]{babel} 
\usepackage[protrusion=true,expansion=true]{microtype} 
\usepackage{amsmath,amsfonts,amsthm}
\usepackage{dsfont}
\usepackage{amssymb}
\usepackage{chngcntr}
\usepackage[toc,page]{appendix}
\usepackage{textcomp}
\usepackage{url}
\usepackage{siunitx}
\usepackage{comment}
\newcommand{\cmtS}[1]{{\color{blue}{(Simon: #1)}}}
\newcommand{\cmtX}[1]{{\color{orange}{(Xiaojun: #1)}}}
\newcommand{\cmtL}[1]{{\color{cyan}{(Liyan: #1)}}}

\usepackage{array}
\usepackage{booktabs} 
\usepackage{natbib}
\usepackage{setspace}
\usepackage{amsmath}

\usepackage{soul}
\usepackage{multirow}
\usepackage{enumitem}
\usepackage{microtype} 
\usepackage[font = small,labelfont=bf,textfont=it]{subcaption}
\usepackage{xcolor}
\usepackage{tabularx}
\usepackage[font = small,labelfont=bf,textfont=it]{caption} 
\usepackage{footnote}
\usepackage[noend]{algpseudocode}
\usepackage{etoolbox}
\usepackage{tabularx}
\usepackage{authblk}
\usepackage{caption}
\usepackage{afterpage}
\usepackage{indentfirst}
\usepackage{enumitem}
\usepackage[ruled,vlined]{algorithm2e}
\newlist{steps}{enumerate}{1}
\setlist[steps, 1]{label = Step \arabic*:}

\algblock{ParFor}{EndParFor}
\algnewcommand\algorithmicparfor{\textbf{for}}
\algnewcommand\algorithmicpardo{\textbf{do\ parallel}}
\algnewcommand\algorithmicendparfor{\textbf{end\ parallel\ for}}
\algrenewtext{ParFor}[1]{\algorithmicparfor\ #1\ \algorithmicpardo}
\algrenewtext{EndParFor}{\algorithmicendparfor}

\newcommand{\bm}[1]{\mathbf{#1}}
\newtheorem{theorem}{Theorem}

\newtheorem{prop}{Proposition}
\usepackage{graphicx}
\usepackage{natbib} 
\usepackage{url} 
\newcommand{\blind}{0}

\usepackage[ruled,vlined]{algorithm2e}

\begin{document}

\def\spacingset#1{\renewcommand{\baselinestretch}%
{#1}\small\normalsize} \spacingset{1}


\if0\blind
{
  \title{\bf PERCEPT: a new online change-point detection method using topological data analysis}
  \author{Xiaojun Zheng$^{a}$, Simon Mak$^{a}$, Liyan Xie$^{b}$, Yao Xie$^{c}$\vspace{0.2in}\\
    $^{a}$Department of Statistical Science, Duke University\vspace{0.05in}\\
    $^{b}$School of Data Science, The Chinese University of Hong Kong, Shenzhen\vspace{0.05in}\\
    $^{c}$H. Milton Stewart School of Industrial and Systems Engineering (ISyE), Georgia Institute of Technology
  }
  \maketitle
} \fi

\if1\blind
{
  \bigskip
  \bigskip
  \bigskip
  \begin{center}
    {\LARGE\bf PERCEPT: a new online change-point detection method using topological data analysis}
\end{center}
  \medskip
} \fi

\bigskip
\begin{abstract}
Topological data analysis (TDA) provides a set of data analysis tools for extracting embedded topological structures from complex high-dimensional datasets. In recent years, TDA has been a rapidly growing field which has found success in a wide range of applications, including signal processing, neuroscience and network analysis. In these applications, the online detection of changes is of crucial importance, but this can be highly challenging since such changes often occur in a low-dimensional embedding within high-dimensional data streams. We thus propose a new method, called PERsistence diagram-based ChangE-PoinT detection (PERCEPT), which leverages the learned topological structure from TDA to sequentially detect changes. PERCEPT follows two key steps: it first learns the embedded topology as a point cloud via persistence diagrams, then applies a non-parametric monitoring approach for detecting changes in the resulting point cloud distributions. This yields a non-parametric, topology-aware framework which can efficiently detect online changes from high-dimensional data streams. We investigate the effectiveness of PERCEPT over existing methods in a suite of numerical experiments where the data streams have an embedded topological structure. We then demonstrate the usefulness of PERCEPT in two applications in solar flare monitoring and human gesture detection.

\end{abstract}

\noindent%
\vfill

\newpage
\spacingset{2} 
\section{Introduction}
\vspace{-0.06in}
\label{sec:intro}

Topological Data Analysis (TDA) is a thriving field at the intersection of statistics, machine learning, and algebraic topology, which has gained traction in recent years. TDA methods provide a set of tools for studying the shapes of complex high-dimensional datasets, by extracting its underlying low-dimensional geometric structures. TDA has found success in a wide range of applications, including signal processing \citep{perea2015sliding}, computational biology \citep{protein}, time series analysis \citep{ts}, and neuroscience \citep{sizemore}. 

Despite promising developments in recent years, there has been little work on integrating topological structure for sequential change-point detection, which is a fundamental problem in many of the aforementioned applications. Change-point detection here refers to the detection of a possible change in signal distribution over time. Traditional change-point detection methods largely focus on likelihood ratio tests, which presume that observations are independently and identically distributed, both before and after the change \citep{page-biometrica-1954,Siegmund1985,poor-hadj-QCD-book-2008,detectAbruptChange93,tartakovsky2014sequential}. When the pre- and post-change distributions are known, one can apply the cumulative sum (CUSUM) detection rule, which has been proved to be optimal \citep{Lorden1971,mous-astat-1986}. In practice, the post-change distribution is typically unknown, in which case one can sequentially estimate it and construct a generalized likelihood ratio test for change detection \citep{lai-ieeetit-1998,lau2018binning}. 

However, a key limitation with these traditional methods is that they may perform poorly for \textit{high-dimensional} data with \textit{complex} low-dimensional structure. This is because the distributions in such setting would typically be difficult to approximate well using parametric families, and the required density ratio would thus be difficult to estimate as well. One solution in this high-dimensional setting is to first extract the underlying low-dimensional structure from data, and then construct detection statistics based on the extracted information. CUSUM-type detection algorithms were proposed in \cite{xie2020seq_ana} and \cite{jiao2018subspace} to detect changes characterized by an unknown subspace structure in the covariance matrix. These methods work well when the true low-dimensional embedding is precisely the embedding used in the detection procedure, i.e., a linear subspace. In complex problems where the true embedding is nonlinear, however, this model misspecification may result in considerable deterioration in detection performance \citep{molloy2017misspecified}. There is thus a need for a high-dimensional change detection method which can integrate a broader yet realistic framework for modeling low-dimensional structure.


In recent years, the rise of TDA methods suggests that, for many problems, the underlying data may have embedded \textit{topological} structure which can be extracted via TDA. Indeed, in such problems, the extracted topology from TDA often captures intuitive features from high-dimensional data which are interpretable for the practitioner. For example, in computer vision, topological features can represent segmented regions of a 3D shape \citep{segmentation}. Similarly, for time series, periodic signals can be captured by topological features via Taken's Embedding \citep{ts}. However, the integration of such structure for change detection has largely remained unexplored. A recent work \citep{ts_cp} proposed a TDA detection approach for time series data, by converting this to a sequence of so-called Betti numbers \citep{persistenthomology}, which capture the number of $k$-dimensional holes on a topological surface. However, Betti numbers are known to provide a weak summary of topology \citep{ghristbarcodes} and thus may be insensitive to certain topological changes. Another recent work \citep{wasserstein_gel} utilized a richer topological summary called a \textit{persistence diagram} \citep{persistenthomology}, which uniquely captures the topological features of the data at different spatial resolutions (more on this in Section \ref{sec:tda}). With persistence diagrams computed at each point in time, the method then makes use of the Wasserstein distance between diagrams from adjacent times as the detection test statistic. However, there are two limitations with this approach: (i) it relies on the Wasserstein distance, which may not fully capture the topological differences between two persistence diagrams; and (ii) its test statistic relies on information from only the immediate past, which can greatly reduce detection efficiency.





To address this, we propose a new \textit{non-parametric}, \textit{topology-aware} framework called PERsistence diagram-based ChangE-PoinT detection (PERCEPT). As in \cite{wasserstein_gel}, PERCEPT leverages the extracted persistence diagrams (which capture topological features of the data at each time) for change detection. However, instead of using the Wasserstein distance of diagrams from adjacent times, PERCEPT extends a recent non-parametric change detection method \citep{yao_cp} to detect changes directly on the diagram point clouds. This yields two advantages: (i) it offers a distribution-based approach which amplifies changes in topological features; and (ii) its test statistic makes use of data within a past time window, which addresses information loss. We demonstrate the effectiveness of PERCEPT over existing methods in a suite of simulation experiments and in applications to solar flare monitoring and human gesture detection.

The rest of the paper is organized as follows. Section \ref{sec:pre} provides preliminaries on persistent homology and motivations. Section \ref{sec:PERCEPT} outlines the PERCEPT methodology. Section \ref{sec:num} demonstrates the effectiveness of this method in numerical experiments. Section \ref{sec:data} applies the method to solar flare monitoring and human gesture detection.

\vspace{-0.3in}
\section{Preliminaries and Motivation} \label{sec:pre}
\vspace{-0.06in}

We first provide a brief overview of TDA, then discuss two baseline methods, the Hotelling's $T^2$ statistic and the Wasserstein distance approach in \cite{wasserstein_gel}, for high-dimensional change-point detection. We then motivate the proposed PERCEPT method via an application to solar flare detection.

\vspace{-0.22in}
\subsection{TDA Preliminaries}
\label{sec:tda}
\vspace{-0.06in}

\begin{figure}[!t]
\vspace{-.4in}
    \centering
    \includegraphics[width=0.9\textwidth]{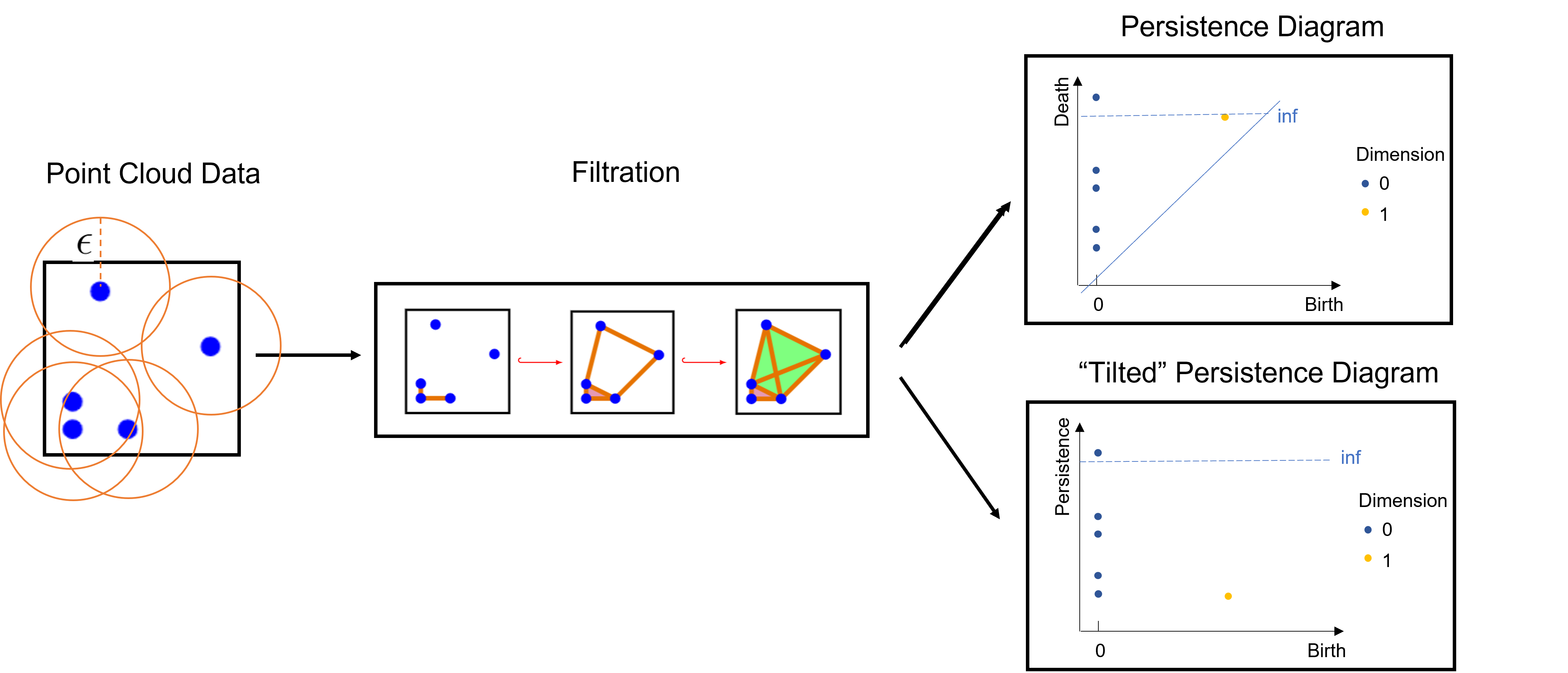}
    \caption{An illustration of the persistent homology pipeline, from point cloud data to a filtration of simplicial complexes to a persistence diagram. The Rips complex with radius $\epsilon$ in the left plot corresponds to the second simplicial complex in the filtration.
   }
    \label{fig:tda}
    
\end{figure}

A primary tool in TDA is \textit{persistent homology}, which extracts topological features (e.g., connected components, holes, and their higher-dimensional analogues) from point cloud data. In what follows, we provide a brief overview of persistent homology, following \cite{ghristbarcodes} and \cite{persistenthomology}. For a given point cloud dataset, persistent homology provides a representation of this as a \textit{simplicial complex}, defined as a set of vertices, edges, triangles, and their higher-dimensional counterparts. A common simplicial complex is the \textit{Rips complex}, which depends on a single scale parameter $\epsilon$. At a given radius $\epsilon > 0$, the Rips complex contains all edges between any two points whose distance is at most $\epsilon$, and contains triangular faces for any three points whose pairwise distances are at most $\epsilon$. Figure \ref{fig:tda} (adapted from \citealp{fig_cite}) illustrates this for a toy dataset. Clearly, a single $\epsilon$ cannot capture all geometric structures embedded in the data. Thus, a sequence of scale parameters is used to build a sequence (or \textit{filtration}) of simplicial complexes. This filtration provides a means for extracting topological structure from the data, e.g., zero-dimensional holes (or connected components), one-dimensional holes, and their higher-dimensional analogues. 

Under this framework, suppose a topological feature appears in the filtration at some radius $\epsilon$ and disappears at a larger radius $\epsilon' > \epsilon$. The pair $(\epsilon, \epsilon')$ gives the \textit{persistence} of the feature, with $\epsilon$ and $\epsilon'$ being its \textit{birth} and \textit{death}, respectively. A prominent topological feature in the point cloud data would have long persistence, whereas a small or noisy topological feature would have short persistence. The persistence information from all topological features can be captured by an (untilted) \textit{persistence diagram} (PD), a collection of points in $\mathbb{R}^2$ where each feature is represented by a point $(\epsilon,\epsilon')$, with $\epsilon$ is its birth time and $\epsilon'$ is its death time. Figure \ref{fig:tda} illustrates the persistent homology pipeline from point cloud data to a persistence diagram. We will distinguish this untilted PD from its tilted variant later. 

\begin{figure}[!t]
\vspace{-.2in}
    \centering
    \includegraphics[width=0.9\textwidth]{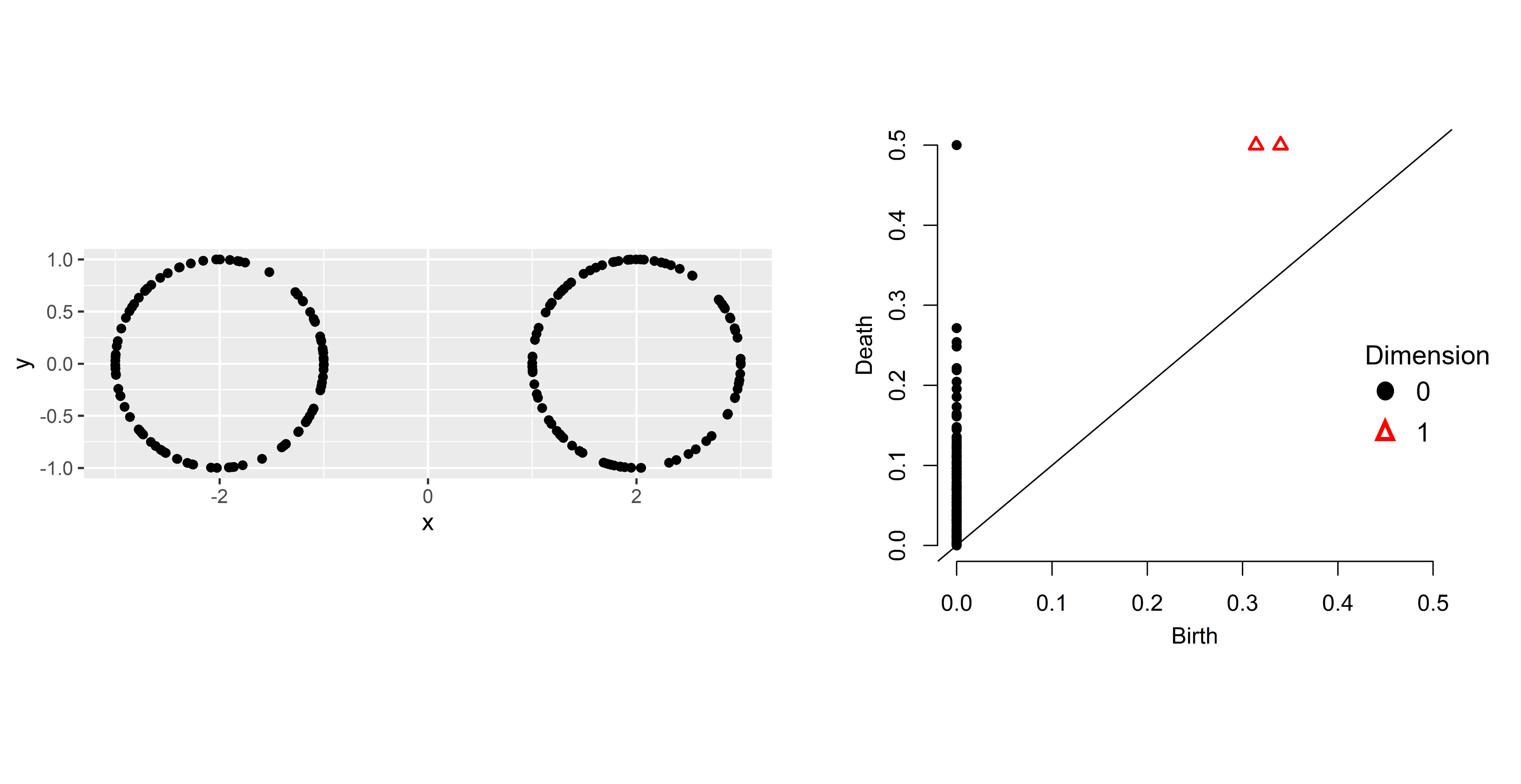}
    \vspace{-0.4in}
    \caption{(left) Point cloud data generated from two disjoint circles, and (right) its corresponding persistence diagram for 0-d and 1-d holes.}
    \label{fig:ph_example}
\end{figure}

We use a simple example to illustrate this translation of topological structure to a PD. Figure \ref{fig:ph_example} (left) shows the point cloud data generated from two disjoint circles with radii $1$, and Figure \ref{fig:ph_example} (right) shows the corresponding PD of this data. For the 0-d holes (black points), we observe many points with small persistence, but one point which persists for a long time. This last point with large persistence suggests the data has two connected components, which is indeed true. For the 1-d holes (red points), we observe two red points with large persistence, which reflects the two holes within the circles.


\vspace{-0.22in}
\subsection{Existing Baseline Methods}
\label{sec:exist}
\vspace{-0.06in}

A classic baseline approach for change detection in multivariate data is the parametric Hotelling's $T^2$ statistic \citep{hotelling1947multivariate}. The Hotelling's $T^2$ statistic makes use of the mean and covariance structure of data, and thus it can detect both the mean and covariance shifts. Given data $\bm{x}_t \in \mathbb{R}^p$ at time $t$, $t = 1, \cdots, T$, the Hotelling's $T^2$ statistic is defined as 
\vspace{-0.2in}
\begin{equation}
    (\bar{\bm{x}} - \boldsymbol{\mu}_0)^\intercal \Sigma_0^{-1} (\bar{\bm{x}} - \boldsymbol{\mu}_0),
\end{equation}

\vspace{-0.2in}
\noindent where $\bar{\bm{x}}$ is the sample mean vector, $\boldsymbol{\mu}_0$ and $\Sigma_0$ is the nominal mean vector and covariance matrix, respectively (this is typically given or estimated from reference data). The vanilla Hotelling's $T^2$ statistic is calculated using \textit{only} data at the current time $t$, with all historical data discarded.

To compute the test statistic more efficiently, it can be coupled with the CUSUM scheme \citep{page-biometrica-1954}, which makes use of a cumulative sum of the statistic over time. The resulting detection statistic $S_t^H$ is then given by: 

\vspace{-0.2in}
\begin{equation}
S_t^{H} = (S_{t-1}^{H})^{+} + (\bar{\boldsymbol{\mu}}_{t-w,t} - \hat{\boldsymbol{\mu}}_0 )^\intercal \widehat{\Sigma}_0^{-1} (\bar{\boldsymbol{\mu}}_{t-w,t} - \hat{\boldsymbol{\mu}}_0 ) - d^{H}, \quad S_0^{H} = 0,
    \label{eq:hotelling}
\end{equation}

\vspace{-0.1in}
\noindent where $(x)^+ = \max\{x, 0\}$, $\bar{\boldsymbol{\mu}}_{t-w,t}$ denotes the sample average of the data vector $\{\bm{x}_{t-w},\ldots,\bm{x}_{t}\}$, and $\hat{\boldsymbol{\mu}}_0$ and $\widehat{\Sigma}_0$ are the \textit{pre-change} mean vector and covariance matrix estimated from historical data. Here, $d^{H}$ is a drift parameter that can also be trained using historical data. When the data is known to be concentrated on a linear subspace, one can adapt the Hotelling's $T^2$ test by first performing Principal Component Analysis (PCA), then using the resulting principal components as data within equation \eqref{eq:hotelling}. With this, a change-point is then declared at time $t$ if the statistic $S_t^{H}$ exceeds a pre-specified threshold $b$.

The second baseline method, the Wasserstein distance approach in \cite{wasserstein_gel}, integrates topology in the following way. The Wasserstein distance (of order 1) for two distributions $P$ and $Q$ on sample space $\Omega$ is defined as:
\vspace{-0.2in}
\[
\mathcal W_1(P,Q) := \min_{\gamma\in\Gamma(P,Q)} \left\{\mathbb E_{(\omega,\omega')\sim\gamma} \left[c(\omega,\omega')\right]  \right\}.
\]  
\noindent Here, $c(\cdot,\cdot)$ is a metric on $\Omega$ (in our case, this is taken as the Euclidean norm), and $\Gamma(P,Q)$ denotes the collection of all (Borel) probability measures on $\Omega\times\Omega$ with marginal distributions $P$ and $Q$.

With this, the Wasserstein distance method is straight-forward. First, at each time $t$, the persistence diagram $\mathcal{D}_t \in \mathbb{R}^2$ is computed from the data $\bm{x}_t$. Next, the Wasserstein distances $S_t^{W} = \mathcal W_1(\mathcal{D}_t,\mathcal{D}_{t+1})$ are computed between the PDs from adjacent time frames. One then declares a change-point at time $t$ when the test statistic $S_t^{W}$ exceeds some pre-specified threshold $b$.

\begin{figure}[!t]
\vspace{-0.4in}
     \centering
     \begin{tabular}{c}
     \includegraphics[width=.9\textwidth]{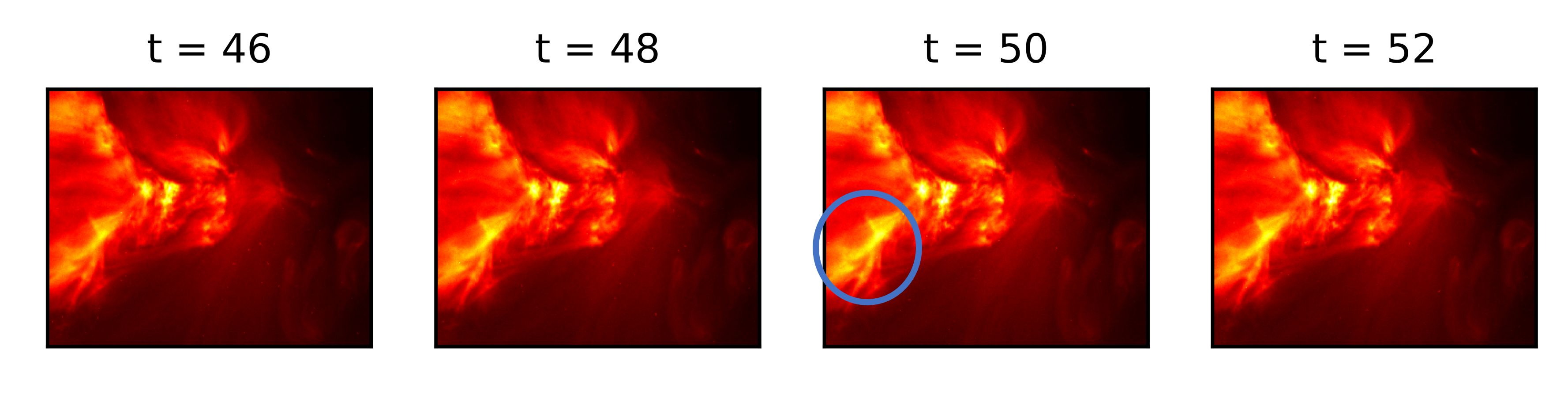}  \\
     (a)
     \end{tabular}
     \vspace{0.1in}
     \begin{tabular}{cc}
         \includegraphics[width=.4\textwidth]{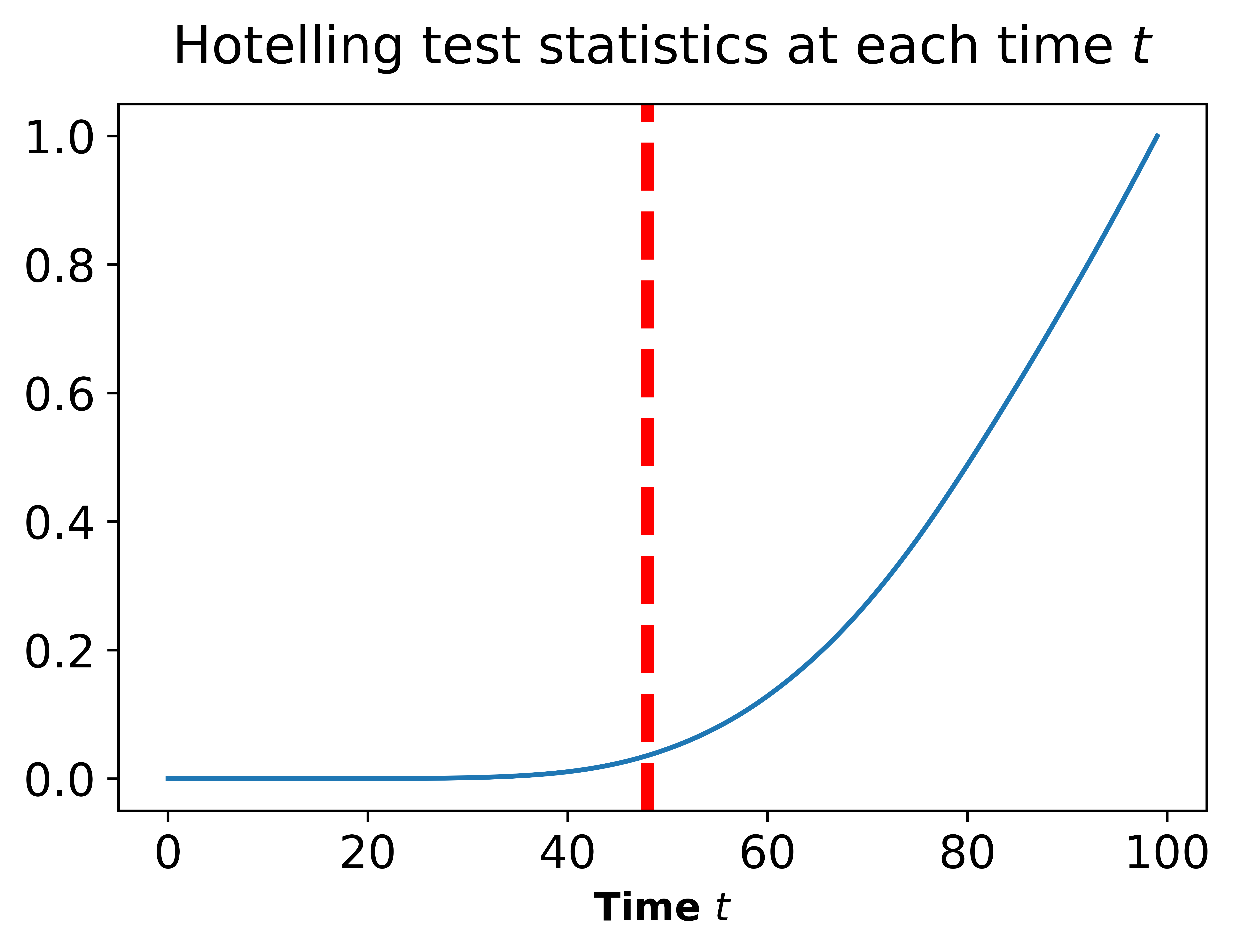} &
         \includegraphics[width=.4\textwidth]{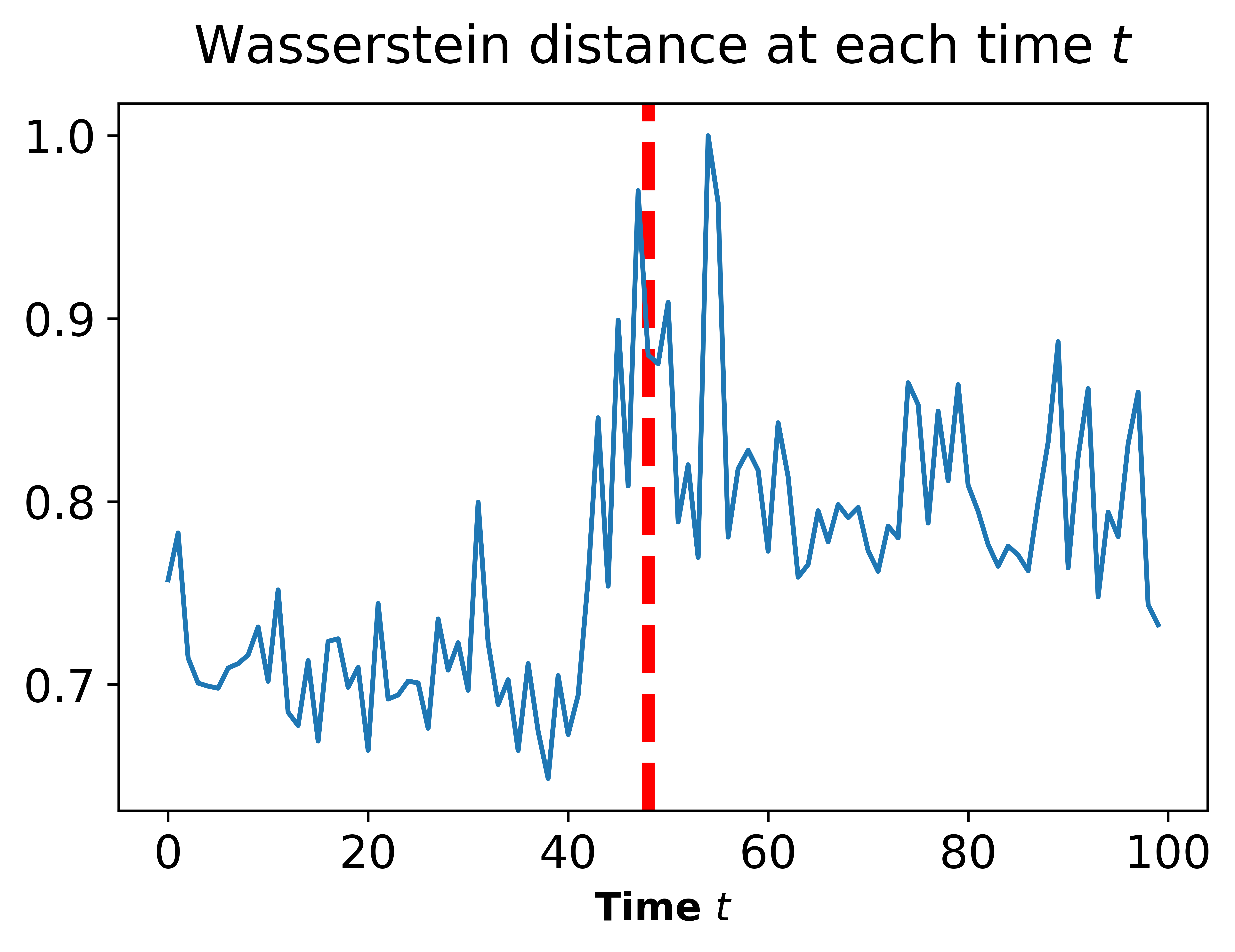} \\
        (b) & (c)
     \end{tabular}
     \caption{(a) Snapshots of the solar flare around the change-point $t^*=49$. (b) Hotelling's $T^2$ statistics at each time $t$. (c) Wasserstein distance statistics at each time $t$.}
     \vspace{-0.1in}
        \label{fig:motivation}
\end{figure}

We investigate these two baseline methods for a motivating solar flare monitoring application (more on this in Section \ref{sec:data}). The data consists of image snapshots ($232 \times 292$ pixels) of a solar flare, captured by the Solar Dynamics Observatory\footnote{See \url{https://sdo.gsfc.nasa.gov/assets/docs/SDO_Guide.pdf}.} at NASA. Figure \ref{fig:motivation}a shows several snapshots of the flare before and after the true change-point at $t^*=49$, where the flare bursts become more pronounced and bright. However, such a change can be quite subtle visually, thus making monitoring a challenging task \citep{yao_solar}. We applied the above two baseline methods, with the Hotelling's $T^2$ applied by first vectorizing the image data, then applying PCA to extract the top 15 principal components. Figures \ref{fig:motivation}b-c show the detection statistics as a function of time $t$. We see that both methods experience a large detection delay from the true change-point at $t^*=49$. For the Hotelling's $T^2$, the test statistic increases quite slowly after the change-point, which suggests it was not capable in capturing the changed image features. Similarly, the Wasserstein distance statistic appears highly unstable and unable to detect the true change-point. Given the limitation of the two baseline methods on solar flare monitoring, we introduce next a new non-parametric, topology-aware method which provides a broader framework for integrating low-dimensional geometric information for monitoring.


\vspace{-0.3in}
\section{Persistence Diagram-based Change  Detection}
\label{sec:PERCEPT}
\vspace{-0.06in}

We now introduce the proposed PERsistence diagram-based ChangE-PoinT (PERCEPT) monitoring method. We first describe the histogram binning for PDs, then show how the extracted histograms can be used for non-parametric change detection. We then present some supporting theory for PERCEPT, and discuss methodological developments on the bin and weight optimization.

\vspace{-0.22in}
\subsection{Persistence Histogram Binning}
\label{sec:perhist}
\vspace{-0.06in}

The first step in PERCEPT is to construct the so-called \textit{persistence histograms}, a novel histogram representation which captures persistence information from the computed PDs. This histogram binning serves two purposes -- it provides a robust way for filtering noise in the PD data, and allows us to leverage recent developments in histogram-based change detection methods within PERCEPT (shown next). To facilitate this, we assume all PDs are given in their \textit{tilted} representation, where a feature is represented by a point $(\epsilon,\epsilon'-\epsilon)$, with $\epsilon$ is its birth time and $\epsilon'-\epsilon$ is its persistence time. Figure \ref{fig:tda} provides an illustration.

Assume, as before, that the PDs $\mathcal{D}_t \in \mathbb{R}^2$ are computed for the data $\bm{x}_t$ at each time $t = 1, \cdots, T$. Further assume that the birth range for the PDs (i.e., the x-axis on $\mathcal{D}_t$) is partitioned into the $L$ bins: $[0,b_1)$, $[b_1,b_2)$, $\ldots$, $[b_{L-1},b_L)$, where $b_l$ is the right break-point for the $l$-th bin. With this, we can now bin the point cloud data $\mathcal{D}_t$. Let $f_{t,l}$ be the sum of persistences for points in $\mathcal{D}_t$ within the $l$-th bin, i.e.:
\vspace{-0.2in}
\begin{equation}
f_{t,l} = \sum_{(u,v) \in \mathcal{D}_t, u \in [b_{l-1},b_l)} v,
\label{eq:w}
\vspace{-0.15in}
\end{equation}
and let $\omega_{t,l} = f_{t,l} / \sum_{l'=1}^L f_{t,l'}$ be its corresponding proportion. The histogram frequencies and distribution of the PD $\mathcal{D}_t$ can thus be represented by the vectors $\bm{f}_t = (f_{t,1}, \cdots, f_{t,L})$ and $\boldsymbol{\omega}_t = (\omega_{t,1}, \cdots, \omega_{t,L})$, respectively. We will call $\boldsymbol{\omega}_t$ the \textit{persistence histogram} of the PD $\mathcal{D}_t$. Figure \ref{fig:hist_cp}a visualizes this binning procedure.

The breakpoints for the persistence histogram (i.e., $b_1, \cdots, b_L$) can be placed such that the bins have equal widths in the birth range using the ``pre-change'' PDs (i.e., the diagrams prior to the change), then kept the same throughout the procedure. Figure \ref{fig:hist_cp}a visualizes this for the aforementioned solar flare application. The left figure shows the binned histogram (with $L=10$ bins) of a PD computed from a pre-change flare image, and the right figure shows the histogram for a ``post-change'' image using the same bins. Given a significant change in topological structure, the corresponding pre-change and post-change persistence histograms (which capture topology information) should be sufficiently different to capture this change. We will leverage this fact to formulate the following test statistic.



\vspace{-0.22in}
\subsection{Test Statistic}
\vspace{-0.06in}

\begin{figure}[!t]
\vspace{-.2in}
    \centering
    \begin{tabular}{cc}
         \includegraphics[width=.58\textwidth]{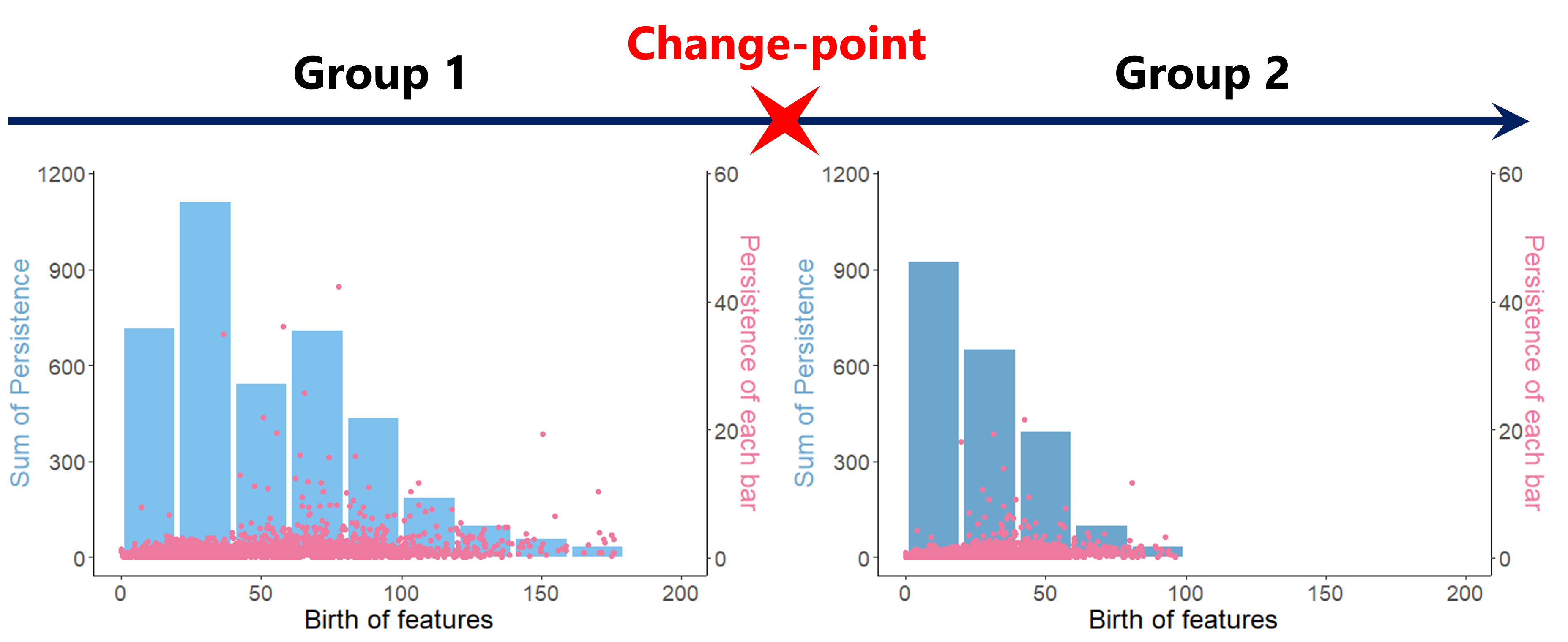}  & \includegraphics[width=.4\textwidth]{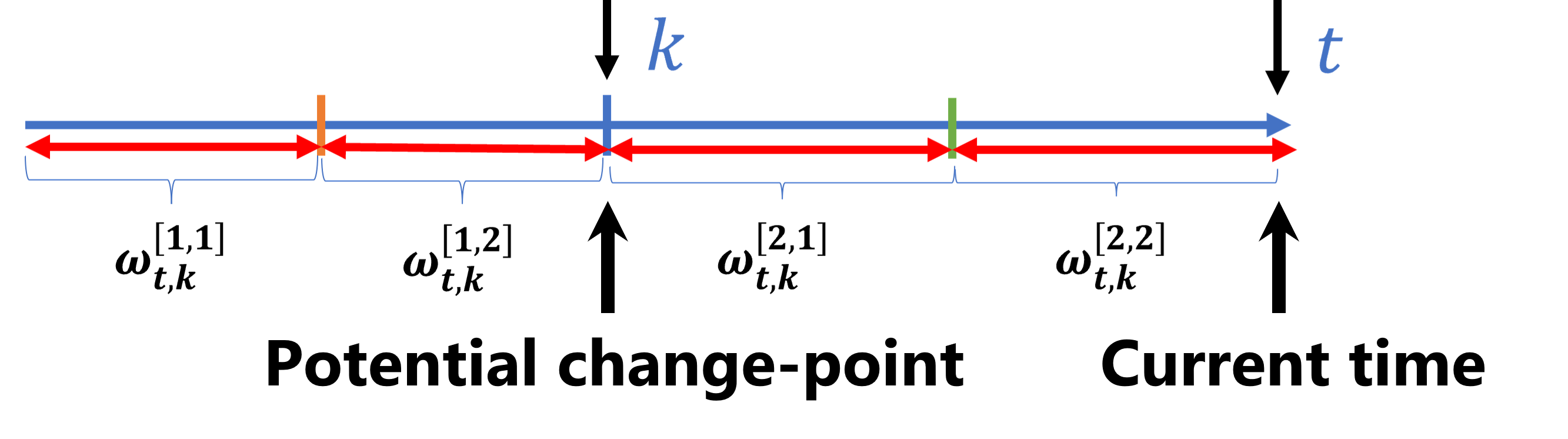} \\
        (a)  & (b)
     \end{tabular}
    \caption{(a) Persistence histograms for a pre-change and post-change solar flare image. (b) Visualizing the intervals used for the weighted $\ell_2$ detection statistic.}
    \label{fig:hist_cp}
    \vspace{-0.15in}
\end{figure}

The second step in PERCEPT is to use the extracted persistence histograms to construct a monitoring test statistic. The idea is as follows. Suppose the bins $[0,b_1), [b_1,b_2)$, $\cdots$, $[b_{L-1},b_L)$ are fixed beforehand. Then, at each time $t$, one can treat the observed persistence histograms $\bm{f}_t$ as \textit{data} sampled from an underlying discrete distribution with $L$ levels. Let $p_{\rm pre}$ denote this so-called \textit{persistence distribution} (with $L$ levels) prior to the change, and $p_{\rm post}$ be the persistence distribution after the change. The goal of detecting topological changes can then be thought of as testing for differences between the persistence distributions $p_{\rm pre}$ and $p_{\rm post}$. We thus desire a test which investigates the following hypotheses:
\vspace{-0.2in}
\begin{equation}
H_0 : p_{\rm pre} = p_{\rm post}, \quad H_A : p_{\rm pre} \neq p_{\rm post}.
\label{eq:hypo}
\vspace{-0.2in}
\end{equation}
To test \eqref{eq:hypo}, we extend a recent non-parametric test in \cite{yao_cp} for detecting changes on discrete distributions. At the current time $t$, we search for all possible change points at time $k < t$ within a fixed-sized window. To investigate whether time $k$ is a change point, we will consider four consecutive time intervals (see Figure \ref{fig:hist_cp}b): the first two intervals are immediately before time $k$, and the last two intervals are immediately after $k$. All four intervals have the same time length of $\lfloor(t-k)/2\rfloor$. We call the first group of intervals (representing potential \textit{pre}-change times) as ``Group 1'', and the latter group (representing potential \textit{post}-change times) as ``Group 2''. We will discuss the choice of $L$ in Section \ref{sec:weight_opt}.

Suppose the histogram breakpoints $b_1, \cdots, b_L$ are given. Consider now the persistence histogram proportions within the two time intervals in Group 1 and the two intervals in Group 2; we denote this as $\boldsymbol{\omega}_{t,k}^{[1,1]}$, $\boldsymbol{\omega}_{t,k}^{[1,2]}$, $\boldsymbol{\omega}_{t,k}^{[2,1]}$ and $\boldsymbol{\omega}_{t,k}^{[2,2]}$, respectively. This non-parametric weighted-$\ell_2$ statistic can then be defined as
\vspace{-0.2in}
\begin{equation}
\chi_{t,k} = (\boldsymbol{\omega}_{t,k}^{[1,1]} -\boldsymbol{\omega}_{t,k}^{[2,1]})^\intercal \Sigma (\boldsymbol{\omega}_{t,k}^{[1,2]}-\boldsymbol{\omega}_{t,k}^{[2,2]}).
\label{eq:l2stat}
\vspace{-0.2in}
\end{equation}
Here, $\Sigma = \text{diag}\{\sigma_1. \cdots, \sigma_L\}$ is the diagonal weight matrix, where $\sigma_l \geq 0, l=1, \cdots, L$. Note that, if time $k$ were indeed a change point, the resulting test statistic $\chi_{t,k}$ would likely be large, thus providing evidence for a change. A simple choice of $\Sigma$ is the identity matrix, which assumes equal weights over persistence histogram bins. We introduce a strategy for optimizing this weight matrix at the end of the section.

With this in hand, an online detection procedure is then given by the stopping time:
\vspace{-0.2in}
\begin{equation}
    \mathcal{T} = \inf\{t: \chi_t^{\max} \geq b\}, \quad \chi_t^{\max} = \max_{0 \leq k \leq t} \chi_{k,t},
    \label{eq:stoptime}
\vspace{-0.2in}
\end{equation}
where $b$ is a pre-specified threshold parameter. Here, $\mathcal{T}$ is the time at which the procedure raises an alarm indicating a change-point has occurred before time $t$, by taking the maximum statistic $\chi_t^{max}$ over all possible change-points $k < t$. A change is declared if and only if the test statistic $\chi_t^{\max}$ exceeds a certain threshold, indicating the two persistence histograms are sufficiently different to declare a change in topology.

The threshold $b$ in \eqref{eq:stoptime} can be set by controlling the false alarm rate to a desired pre-specified level, or, equivalently, by controlling the average run length (ARL, more on this later) to be above a pre-specified level. For the above weighted-$l_2$ statistic, theoretical approximations are available for the ARL as a function of threshold $b$ \citep{yao_cp}:
\vspace{-0.2in}
\[
\mathbb E_\infty[\mathcal T] = \frac12 \frac{b^{-1} e^{b^2/(2\sigma_p^2) }[2\pi \sigma_p^2]^{1/2}}{ \int_{[4b^2/(m_1\sigma_p^2)]^{1/2}}^{[4b^2/(m_0\sigma_p^2)]^{1/2}} y \nu^2(y)dy}(1+o(1)).
\label{eq:ARL}
\]
Here, $\sigma_p^2 :=  4[ \sum_{l=1}^L \sigma_i^2p_i^2(1-p_i)^2 + \sum_{i\neq j}\sigma_i\sigma_j p_i^2p_j^2]$, where $p :=  p_{\rm pre}$, $m_0$ and $m_1$ are two known constants, and $\nu(\cdot)$ is a special function that is easy to calculate. Further details can be found in \cite{yao_cp}. With this approximation, one can then set the threshold $b$ to bound the ARL $\mathbb E_\infty[\mathcal T]$, which then controls the false alarm rate.

\vspace{-0.22in}
\subsection{Connection between EDD and Topology}
\label{sec:theory}
\vspace{-0.06in}




We now provide an interesting connection between the expected detection delay (EDD) and topology, which sheds light on how PERCEPT may be useful for topological change detection. Consider the following definitions of EDD and Average Run Length (ARL), two fundamental metrics in online change-point detection. Let $\mathbb{E}_\infty$ denote the expectation under the probability measure when there is no change-point (i.e., the change-point equals to $\infty$), and let $\mathbb{E}_0$ denote the expectation under the probability measure when the change happens at time 0. For a given stopping time $\mathcal{T}$ of a monitoring procedure, its ARL is defined as $\mathbb{E}_\infty[\mathcal{T}]$, the expected run length to false alarm when there is no change, and the EDD is defined as $\mathbb{E}_0[\mathcal{T}]$, the number of samples needed to detect the change. Theoretically, the EDD is known to be linearly related to the $\log$(ARL) \citep{tartakovsky2014sequential}.


Next, we introduce the bottleneck distance, a standard metric for topological distance \citep{ghristbarcodes}. Suppose, for two point cloud datasets with different topologies, one computes its corresponding (untilted) PDs $\mathcal{D}_1$ and $\mathcal{D}_2$, respectively. The extracted topological features in these PDs can then be compared via a \textit{matching} $\eta$. This matching is performed in two steps: (i) it pairs each point in the first PD $\mathcal{D}_1$ with a point in the second PD $\mathcal{D}_2$ or a point on the diagonal line, and (ii) it pairs each point in $\mathcal{D}_2$ with a point in $\mathcal{D}_1$ or a point on the diagonal. The \textit{bottleneck distance} \citep{ghristbarcodes} between the PDs $\mathcal{D}_1$ and $\mathcal{D}_2$ is then defined as:
 \[ d_B(\mathcal{D}_1, \mathcal{D}_2) = \inf_\eta \sup_{\bm{y} \in \mathcal{D}_1} || \bm{y} - \eta(\bm{y}) ||_\infty. \]
Here, the supremum is taken over all matched points in $\mathcal{D}_1$, and the infimum is taken over all possible matchings $\eta$. Clearly, a larger bottleneck distance indicates the extracted features from the first PD are quite different from that for the second PD (and vice versa). This then suggests the topology for the first dataset is markedly different from that for the second dataset. This link between the bottleneck distance and topological differences is formalized by the Stability Theorem \citep{stability}, a key theorem in TDA.

With this in hand, the EDD of the proposed PERCEPT method can be then linked to the bottleneck distance of the topologies of the pre- and post-change data. Recall that PERCEPT makes use of the non-parametric $l_2$-statistics in \cite{yao_cp} on the underlying persistence diagrams. It is known \citep{yao_cp} that the EDD for such a procedure can be upper bounded by: \begin{equation}
\text{EDD} \leq \frac{2b}{({\min_{i} \Sigma_{ii}) ||p_{\rm pre}-p_{\rm post}||_2^2}},
\label{eq:l2edd}
\end{equation}
where $b$ is the pre-specified detection threshold and $\Sigma_{ii}$ is the $i$-th diagonal entry of $\Sigma$. In other words, the larger the difference between the pre- and post-change persistence distributions $p_{\rm pre}$ and $p_{\rm post}$, the smaller the EDD for PERCEPT. We can then show (technical details in Appendix \ref{sec:prop}) that, under certain asymptotic approximations, this difference $||p_{\rm pre}-p_{\rm post}||_2^2$ can be lower bounded by the bottleneck distance between a pre-change PD $\mathcal{D}_{\rm pre}$ and post-change PD $\mathcal{D}_{\rm post}$, which quantifies differences in pre- and post-change topology. Hence, together with \eqref{eq:l2edd}, this suggests that, for PERCEPT, \textit{the greater (or smaller) the topological difference is between pre- and post-change data, the smaller (or greater) its detection delay}, which is as desired.

\vspace{-0.22in}
\subsection{Persistence Cluster Binning}
\label{sec:voronoi}
\vspace{-0.06in}

The persistence histogram binning approach (Section \ref{sec:perhist}) can be viewed as partitioning the persistence space into vertical \textit{rectangular} regions, which are then used to bin the persistence diagrams for monitoring. This may have two practical limitations. First, since persistences are summed within each bin, the procedure can distinguish topological features with different birth times, but not features with similar birth times but different persistences. Second, the restriction of partitions to be vertical and rectangular may hamper the ability of this method in identifying regions of greatest change between the pre- and post-change distributions. To address this, we propose an alternate novel \textit{persistence clustering} approach, which provides a more general partition of the persistence space.

\begin{figure}[!t]
\vspace{-0.7in}
    \centering
    \begin{tabular}{cc}
       \includegraphics[width=0.48\textwidth]{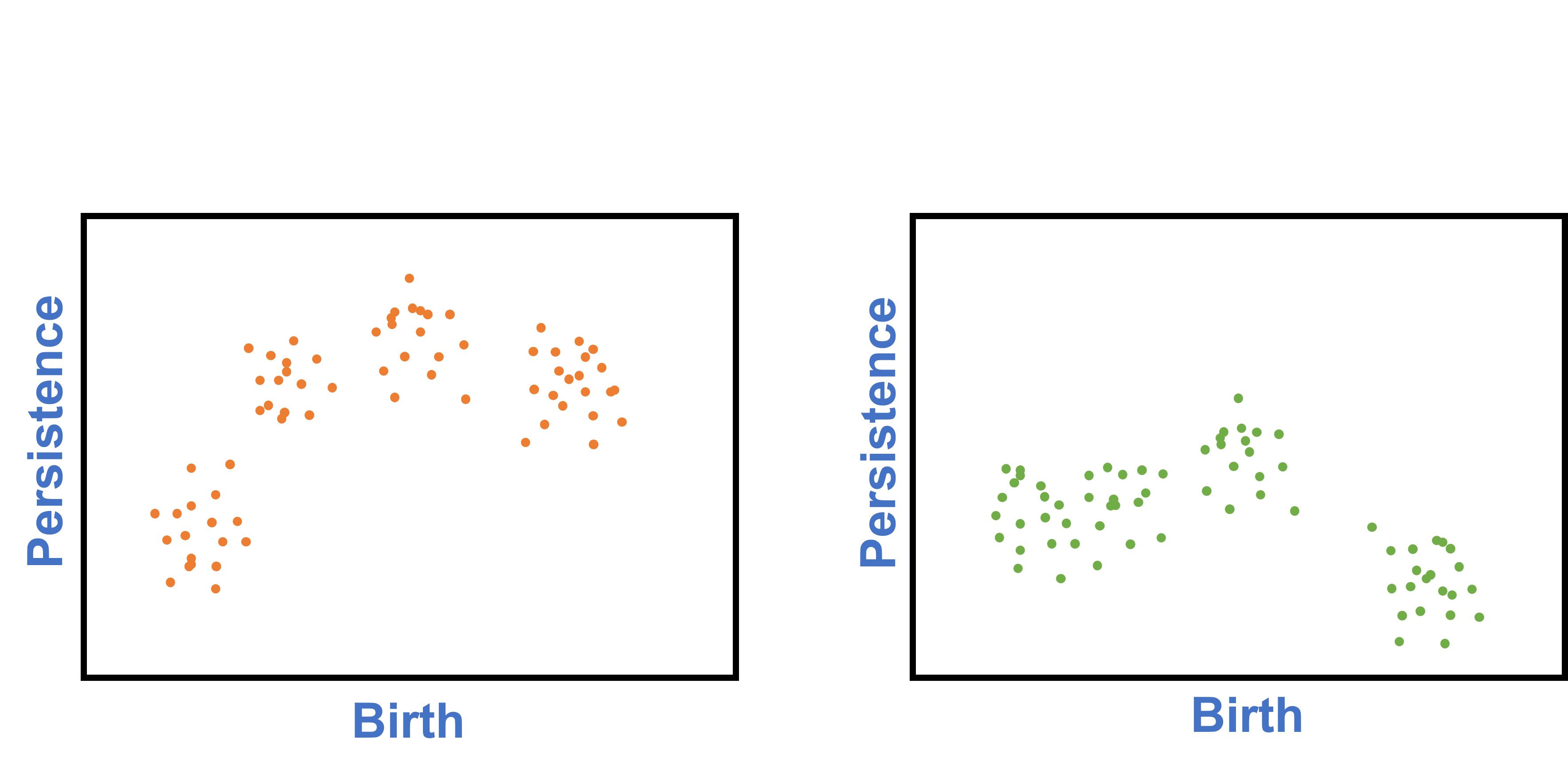}&
       \includegraphics[width=0.48\textwidth]{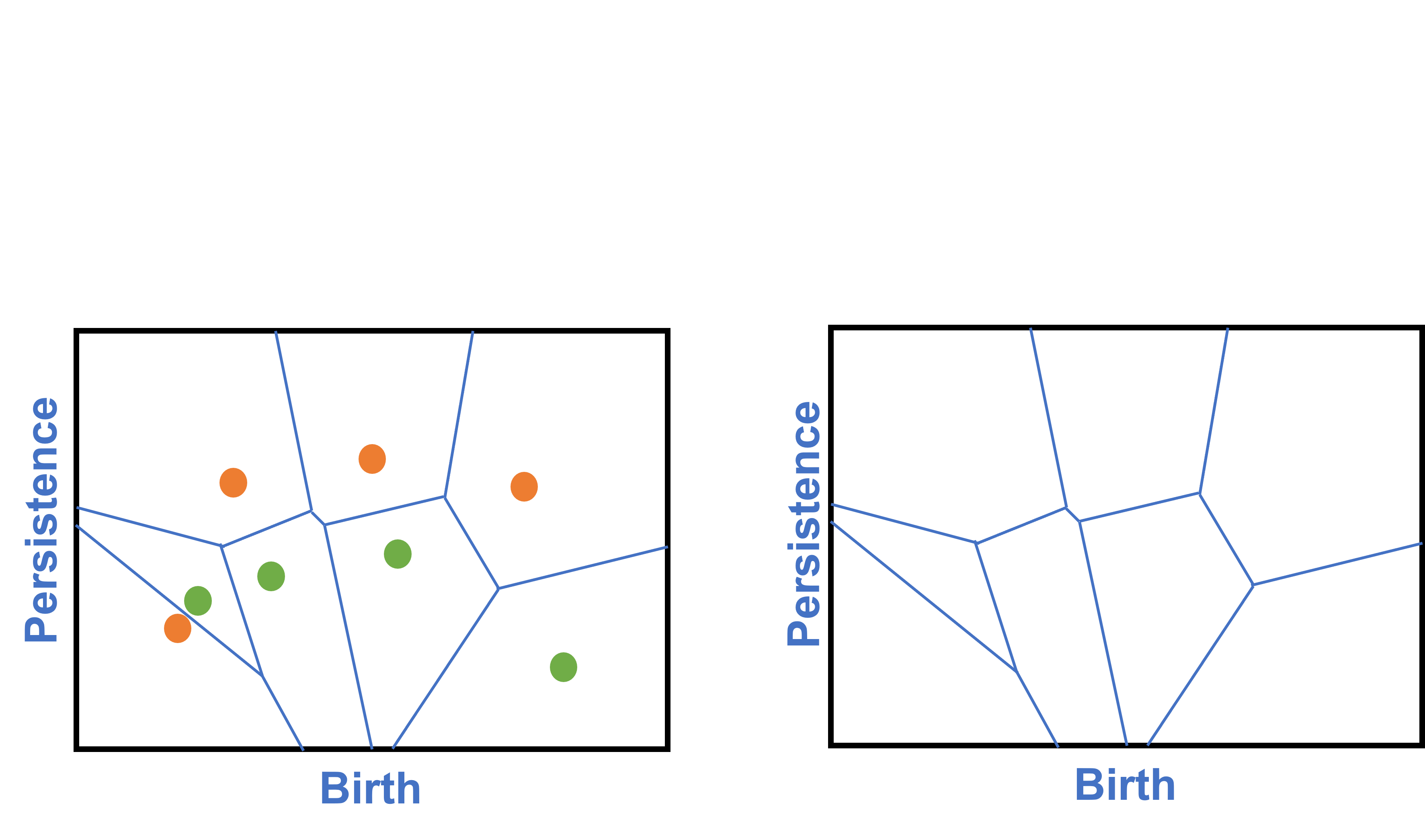}\\
       (a) & (b)\\
    \end{tabular}
    \caption{(a) A sample pre-change (orange) and post-change (green) PD. (b) Pre- and post-change cluster centers and its corresponding persistence cluster via Voronoi diagrams.}
    \label{voronoi}
\end{figure}

Suppose we have training PDs for both pre- and post-change regimes. The idea is to find a clustering of these point clouds, so that the corresponding partition of the persistence space can discriminate well the pre- and post-change distributions. We construct these \textit{persistence clusters} as follows. First, we perform $k$-means clustering \citep{lloyd1982least} on the pre-change PD point clouds, which returns $C_{\rm pre}$ cluster centers $\mathcal{C}_{\rm pre} = \bm{c}_1, \cdots, \bm{c}_{C_{\rm pre}}$. We can do the same clustering on the post-change PDs to obtain $C_{\rm post}$ cluster centers $\mathcal{C}_{\rm post} = \bm{c}_1', \cdots, \bm{c}_{C_{\rm post}}'$. Using both the pre- and post-change centers $\mathcal{C} = \mathcal{C}_{\rm pre} \cup \mathcal{C}_{\rm post}$, we then form a \textit{Voronoi diagram} using centers $\mathcal{C}$ \citep{aurenhammer1991voronoi}, i.e., a partition of $\mathbb{R}^2$ to its closest point in $\mathcal{C}$. Figure \ref{voronoi} visualizes this persistence clustering procedure. The number of clusters can be determined by the elbow method in $k$-means clustering \citep{elbow}. 

With this, we then employ the same weighted-$\ell_2$ test statistic in \eqref{eq:l2stat}. The only difference is that, instead of taking $\boldsymbol{\omega}_{t,k}^{[1,1]}$, $\boldsymbol{\omega}_{t,k}^{[1,2]}$, $\boldsymbol{\omega}_{t,k}^{[2,1]}$ and $\boldsymbol{\omega}_{t,k}^{[2,2]}$ as persistence histogram proportions, we take these as the proportions over persistence clusters. As Figure \ref{voronoi}b shows, these persistence clusters may yield improved discrimination between the pre- and post-change distributions (compared to the earlier persistence histograms), particularly when there is a large amount of points (or features) captured in the persistence diagrams. When there is only a small number of features, however, the earlier persistence histogram approach should be used instead, since there may be insufficient data to fit the more complex persistence clusters. We will demonstrate this later in numerical experiments.





\vspace{-0.22in}
\subsection{Weight Optimization}
\label{sec:weight_opt}
\vspace{-0.06in}

We now discuss the specification of the weights $\Sigma = \text{diag}\{\sigma_1, \cdots, \sigma_L\}$ in the test statistic \eqref{eq:l2stat}. When no training data are available on both pre- and post-change regimes, a standard approach would be to set unit weights over all bins, i.e., $\sigma_l = 1$. However, when such training data are available, we may be able to find a non-uniform choice of weights which amplifies the differences between the pre- and post-change distributions. In our implementation later, we adopt the weight optimization approach in \cite{yao_cp}, which aims to find weights $\Sigma$ to maximize the worst-case weighted $\ell_2$ distance. This can be formulated as:
\vspace{-.2in}
\begin{equation}
\begin{gathered}
    \max_{\sigma \geq 0, g(\sigma) \leq 1}{f(\sigma)}, \\
    f(\sigma) := \min_{p_{\rm pre},p_{\rm post}} \left\{ \sum_{i} \sigma_i (p_{{\rm pre},i}-p_{{\rm post},i})^2: p_{\rm pre},p_{\rm post} \in \Delta, ||p_{\rm pre}-p_{\rm post}||_2 \geq \rho\right\}\\
    g(\sigma) := \max_{p_{\rm pre} \in \Delta} \sum_{i} \sigma_i^2p_{{\rm pre},i}^2
    \label{eq:weightopt}
    \end{gathered}
\end{equation}
Here $p_{\rm pre,i}$ and $p_{\rm post,i}$ denote the $i$-th entry in the vector $p_{\rm pre}$ and $p_{\rm post}$. The minimization in $f(\sigma)$ is taken over all possible pre- and post-change distributions $p_{\rm pre}$ and $p_{\rm post}$, within the probability simplex $\Delta$, that are $\rho$-separable for a given $\rho > 0$. The choice of $\rho$ depends on the scale of change that one wants to detect; in our experiments later, $\rho$ is set to be $0.1$. For persistence histograms, the optimal number of bins $L$ can be optimized simultaneously by selecting $L$ which yields the highest $f(\sigma)$ in \eqref{eq:weightopt}. In practice, the training pre- and post-change data are required for estimating the persistence distributions $p_{\rm pre}$ and $p_{\rm post}$. Further details on this optimization can be found in \cite{yao_cp}. These optimized weights can then be used within the test statistic \eqref{eq:l2stat} for change detection.

In implementation, we have found that a small modification of \eqref{eq:weightopt} can yield noticeably improved performance. Note that when $p_{\rm pre,i}/p_{\rm post,i}$ or $p_{\rm post,i}/p_{\rm pre,i}$ is exceedingly larger from $1$, and $|p_{\rm pre,i}-p_{\rm post,i}|$ is small for some $i$, it may be difficult to pick out bin $i$ as an important bin using the above formulation. However, such a bin $i$ distinguishes the pre- and post-change distributions because their relative difference is large. Thus, we see that this optimization is distribution-dependent; when $p_{\rm pre},p_{\rm post}$ are extremely different from the uniform distribution, the above weights may not be ideal for distribution discrimination. In practice, we suggest using the \textit{relative} difference of $p_{\rm pre}$ and $p_{\rm post}$ instead of the \textit{absolute} difference in the above formulation, i.e., minimizing $\sum_{i} \sigma_i \{(p_{\rm pre,i}-p_{\rm post,i})/p_{\rm post,i}\}^2$ instead of $\sum_{i} \sigma_i (p_{\rm pre,i}-p_{\rm post,i})^2$ for $f(\sigma)$ in \eqref{eq:weightopt}.





\vspace{-0.3in}
\section{Simulation Study}\label{sec:num}
\vspace{-0.06in}

We now explore the performance of the proposed PERCEPT method in a suite of simulation studies where the underlying data is generated with topological structure. We investigate several challenging change scenarios, including topology changes, noise changes, and its scalability for higher-dimensional data.

Three baseline methods are used here for comparison. The first is the aforementioned parametric Hotelling's $T^2$ test, using the 15 extracted principal components from PCA on the original data. The second is the Wasserstein distance method proposed in \cite{wasserstein_gel}, which makes use of the Wasserstein distance between PDs in adjacent times. Details on both methods can be found in Section \ref{sec:intro}. The third method is 
the maximum mean discrepancy (MMD) test \citep{mmd,li2015m}, a widely used non-parametric change detection method. Given a class of functions $\mathcal F$ and two distributions $p$ and $q$, the MMD distance between $p$ and $q$ is defined as 
$\mathrm{MMD}_{\mathcal F}(p,q) = \sup_{f\in\mathcal F}(\mathbb E_{x\sim p}[f(x)] - \mathbb E_{y\sim q}[f(y)])$. 
When $\mathcal{F}$ is a reproducing kernel Hilbert space (RKHS) associated with kernel function $K(\cdot,\cdot)$, this MMD statistic can be written as: 
\begin{equation}
S^M_t = \frac{1}{n_{\rm pre}^2} \sum_{i,i'=1}^{n_{\rm pre}} K(\boldsymbol{\bm{x}}_i, \boldsymbol{\bm{x}}_{i^{'}}) + \frac{1}{n_{\rm post}^2} \sum_{j,j'=1}^{n_{\rm post}} K(\boldsymbol{\bm{x}}_j, \boldsymbol{\bm{x}}_{j^{'}}) - \frac{2}{n_{\rm pre} n_{\rm post}} \sum_{i=1}^{n_{\rm pre}}\sum_{j=1}^{n_{\rm post}}K(\boldsymbol{\bm{x}}_i, \boldsymbol{\bm{x}}_j).
\label{eq:mmd}
\end{equation}
In our implementation, we used the standard Gaussian radius basis function (RBF) kernel $K(\cdot,\cdot)$, where the kernel bandwidth is chosen using the so-called ``median trick'' \citep{med_trick}, i.e., set to be the median of the pairwise distances between data points. For computational efficiency, instead of searching for all possible change-point $k<t$, we adopt a window-limited procedure which considers only $k \in [t-m_1, t-m_0]$, where $m_0=20$ and $m_1=80$ throughout the experiments.

The simulation set-up is as follows. We simulate data $\bm{x}_1, \cdots, \bm{x}_T$ with $T = 400$, and the change-point is set at time $t^*=200$. In other words, $\bm{x}_1, \cdots, \bm{x}_{200}$ are generated from the pre-change distribution, and $\bm{x}_{201}, \cdots, \bm{x}_{400}$ are generated from the post-change distribution. In our simulations, this point cloud data is generated with topological structure from two simple mgeometric shapes, the unit sphere and the ellipsoid (in varying dimensions). Under a quick inspection of the persistence diagrams, we decide to use the persistence clustering approach across all simulation studies. The goal is to have PERCEPT learn this topological structure from data, then make use of such structure to quickly identify change-points. 

\vspace{-0.22in}
\subsection{Shape Change}
\vspace{-0.06in}

\begin{figure}[!t]
\vspace{-0.2in}
    \centering
    \begin{tabular}{cc}
    \hspace{-0.05in}
    \includegraphics[width=0.45\textwidth]{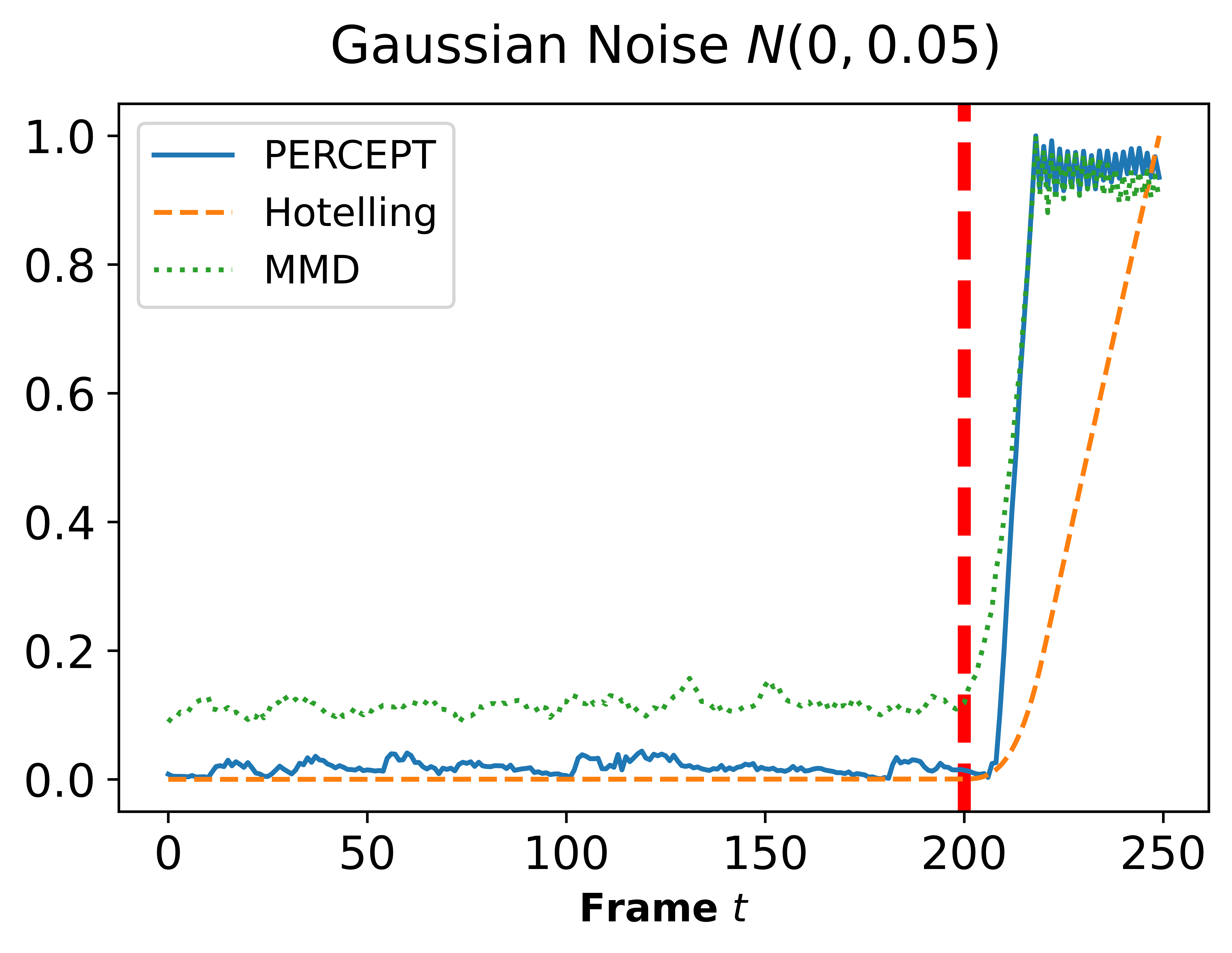}
    \hspace{0.05in}
    \includegraphics[width=0.45\textwidth]{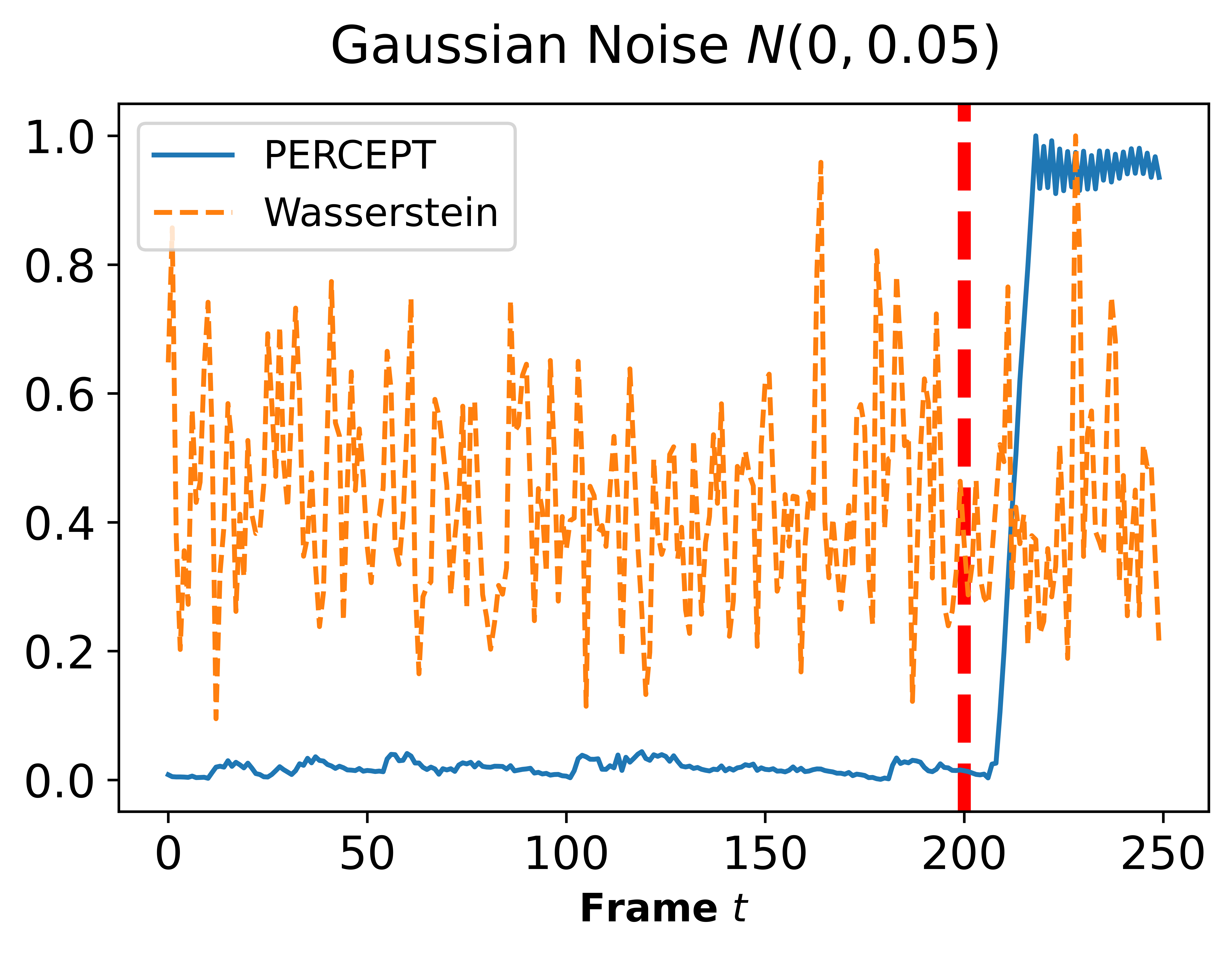}\\
    (a)
    \end{tabular}
    
    \begin{tabular}{cc}
    \includegraphics[width=0.45\textwidth]{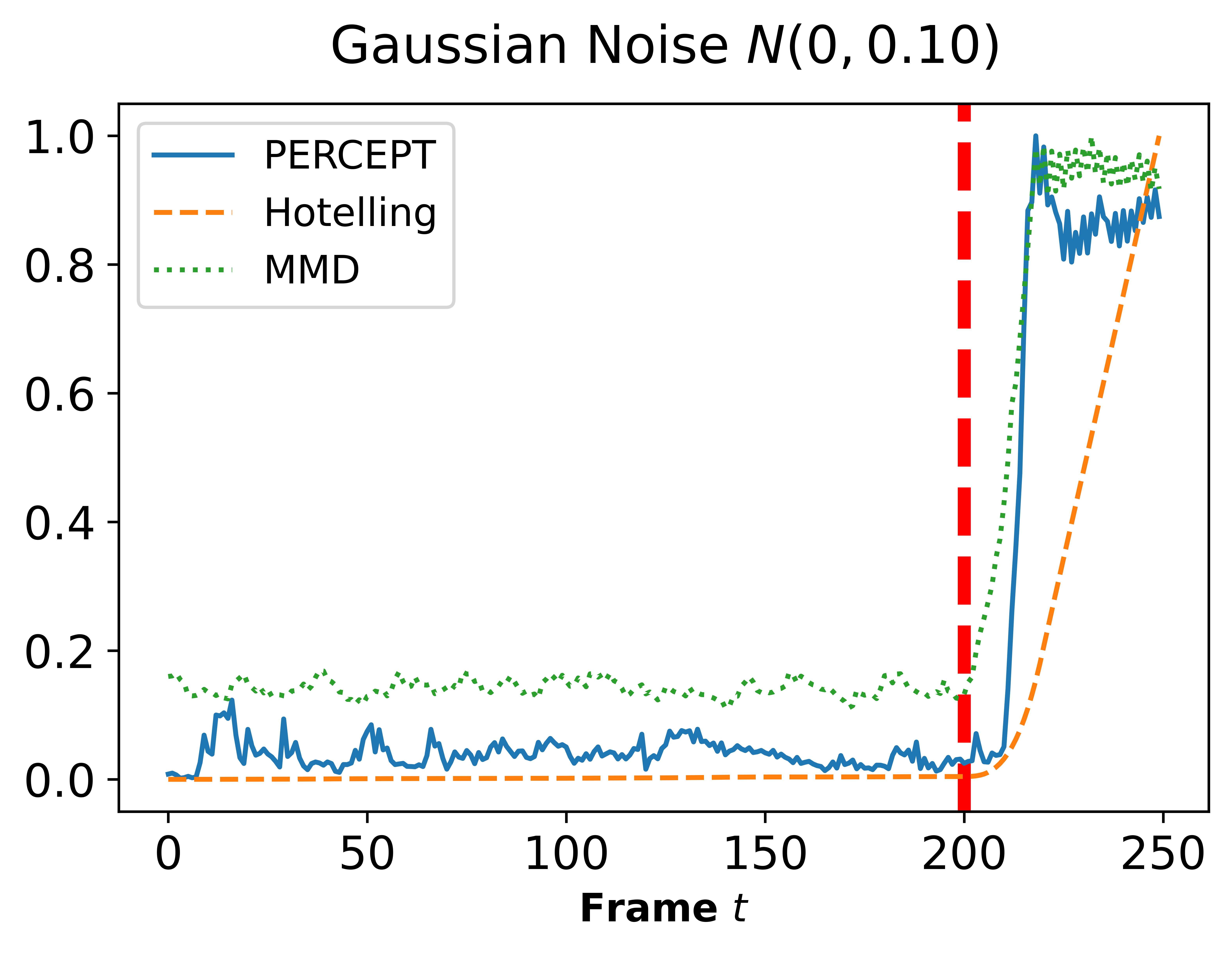} 
    \includegraphics[width=0.45\textwidth]{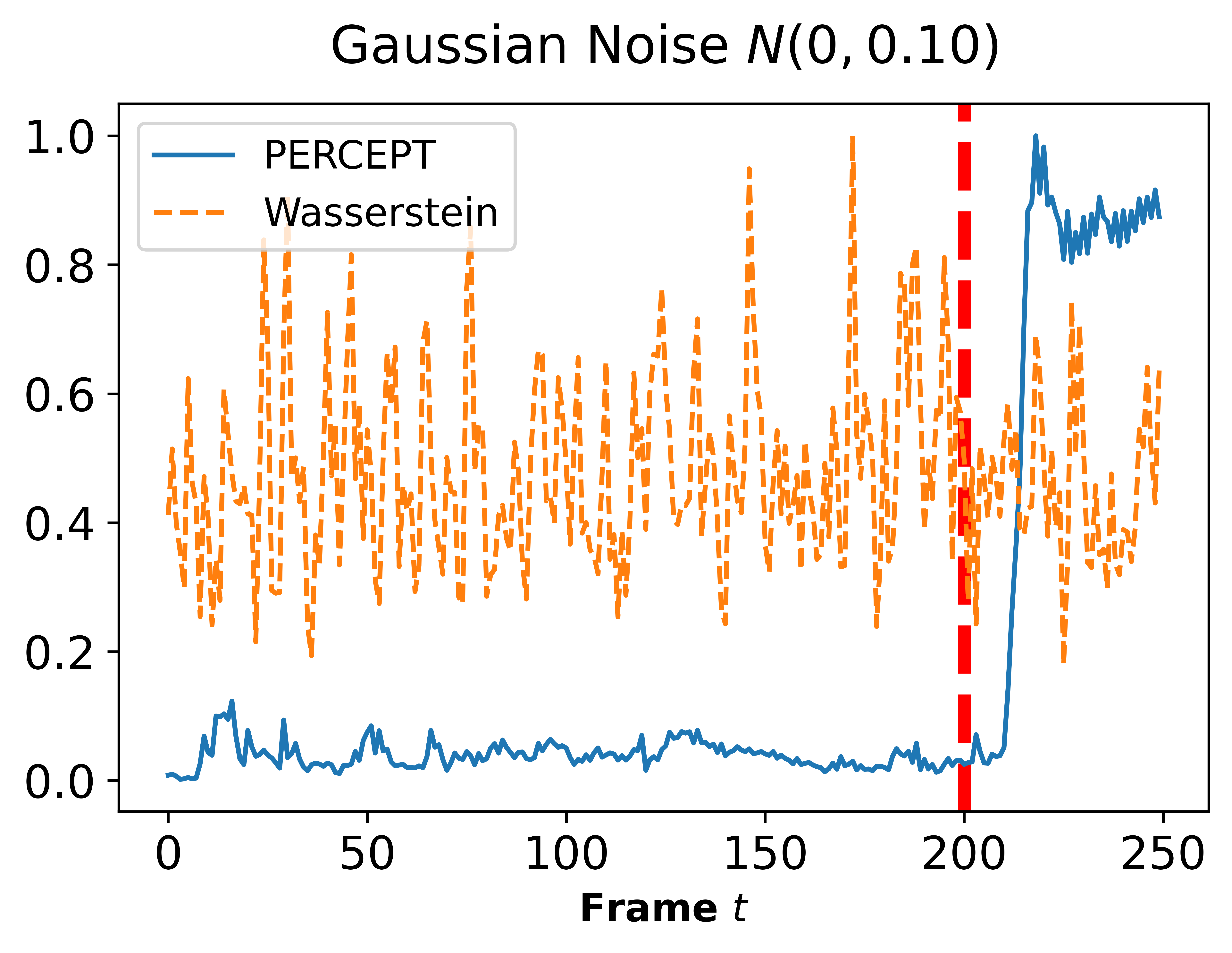}\\
    (b)
    \end{tabular}
     \caption{(a) The $\ell_2$ detection statistic $\chi_t^{\max}$, Hotelling's $T^2$ statistics, MMD statistics and Wasserstein distance at each time $t$ for Gaussian noise $N(0,0.05)$, with the vertical red dashed line indicating the true change-point. (b) Same for Gaussian noise $N(0,0.10)$.}
    \label{shape-change}
\end{figure}

We first consider the case of geometric shape changes, where the pre-change data is sampled with noise from the unit two-dimensional (2-D) circle, and the post-change data is sampled with noise from a 2-D ellipse. Two noise settings are considered for this experiment: $N(0, 0.05)$ and $N(0, 0.10)$. Figure \ref{shape-change} shows the detection statistics from PERCEPT, Hotelling's $T^2$ and MMD under the two noise settings. We see that both PERCEPT and MMD are able to quickly detect the change-point: both monitoring statistics peak up immediately after the change-point at $t=200$. The Hotelling's $T^2$ statistic, however, shows a relatively larger delay. This suggests that, by integrating topological structure, PERCEPT can yield improved performance over the Hotelling's $T^2$. For the Wasserstein distance method, the statistic experiences a large spike at exactly $t=200$ for the noise setting $N(0, 0.05)$, but is unable to detect the change at the larger noise level $N(0, 0.10)$. 


\vspace{-0.22in}
\subsection{Noise Change}\label{sec:numerical_noise}
\vspace{-0.06in}

\begin{figure}[!t]
\vspace{-0.2in}
    \centering
    \begin{tabular}{cc}
    \hspace{-0.05in}
    \includegraphics[width=0.45\textwidth]{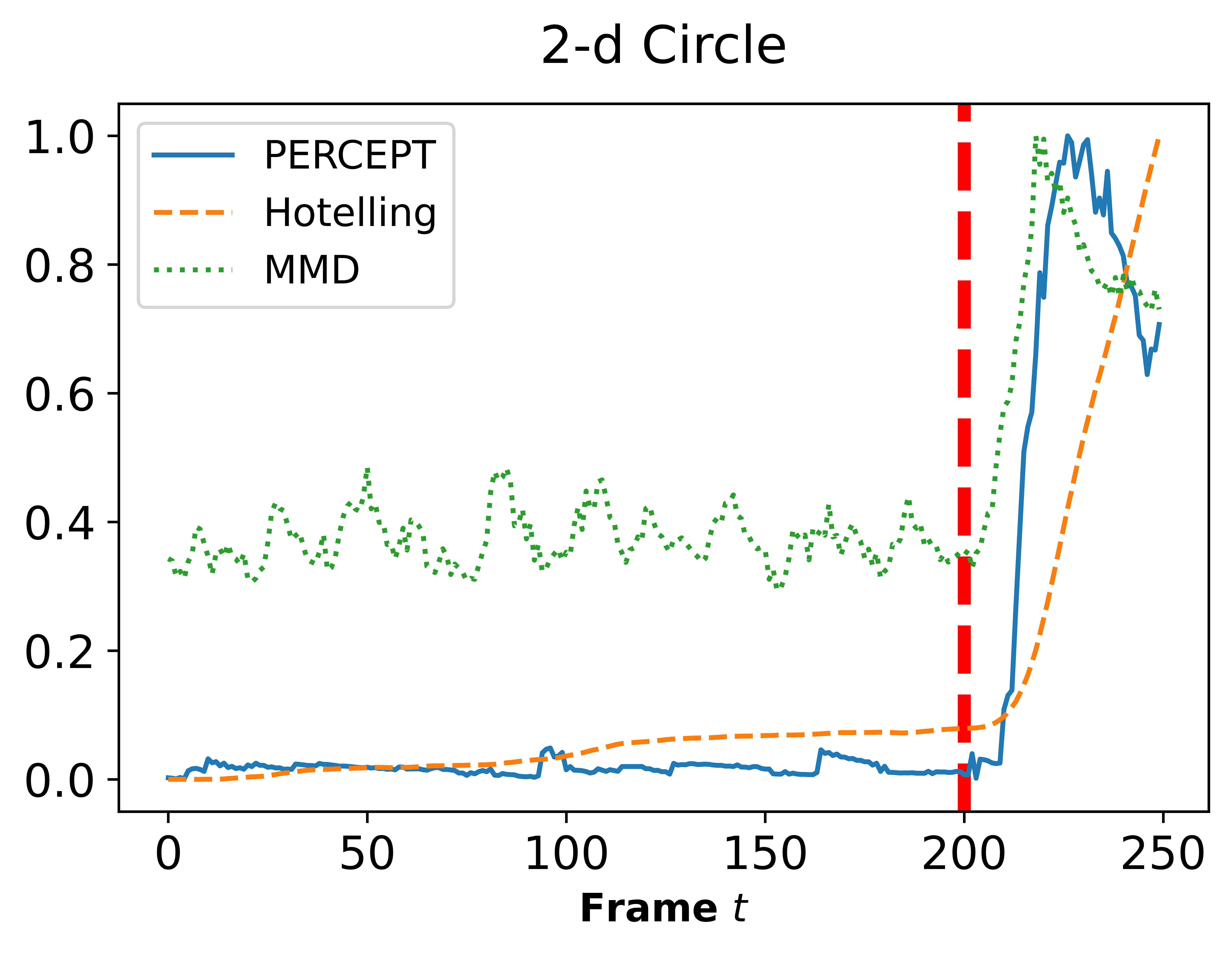}
    \hspace{0.05in}
    \includegraphics[width=0.45\textwidth]{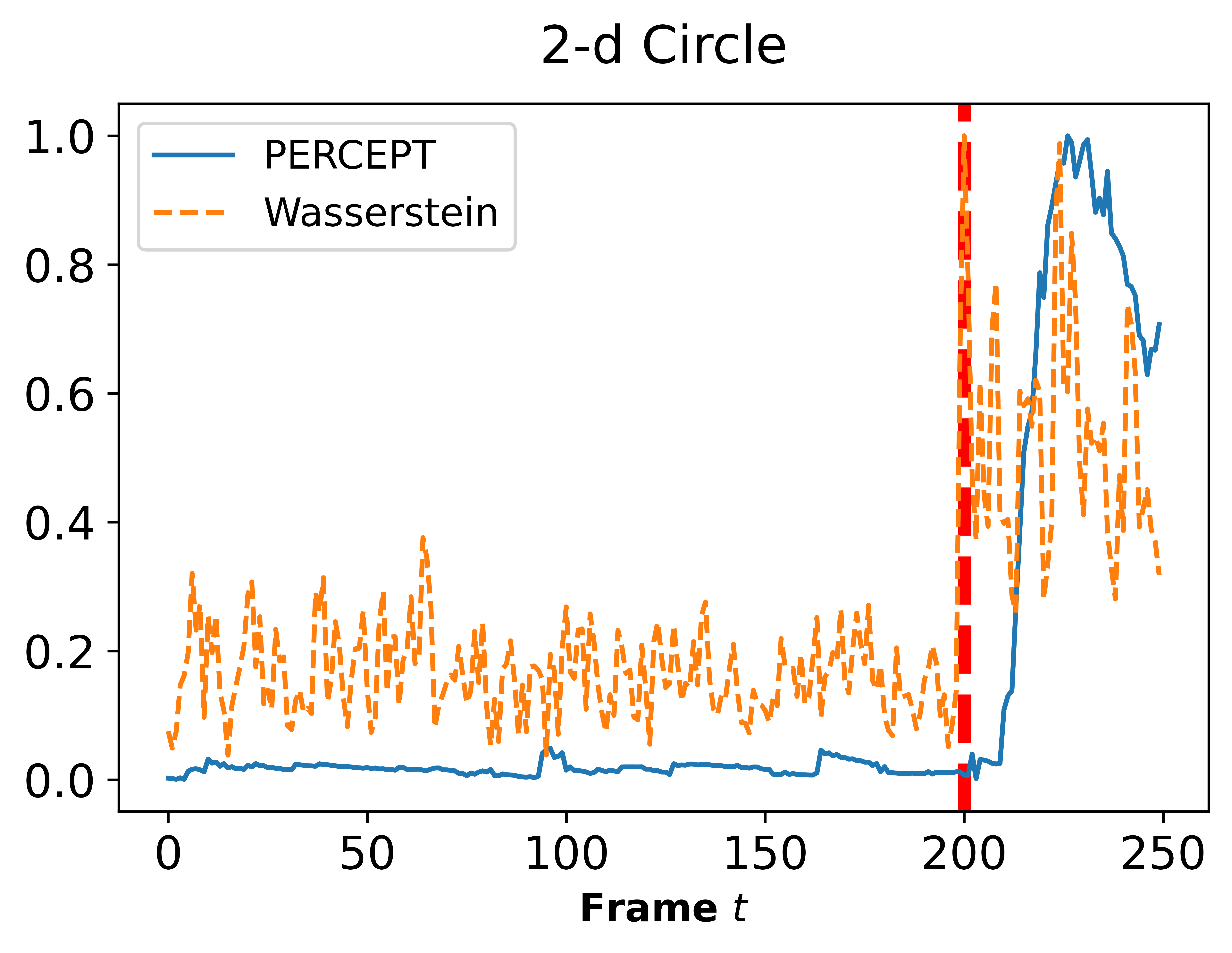}\\
    (a)
    \end{tabular}
    
    \begin{tabular}{cc}
    \includegraphics[width=0.45\textwidth]{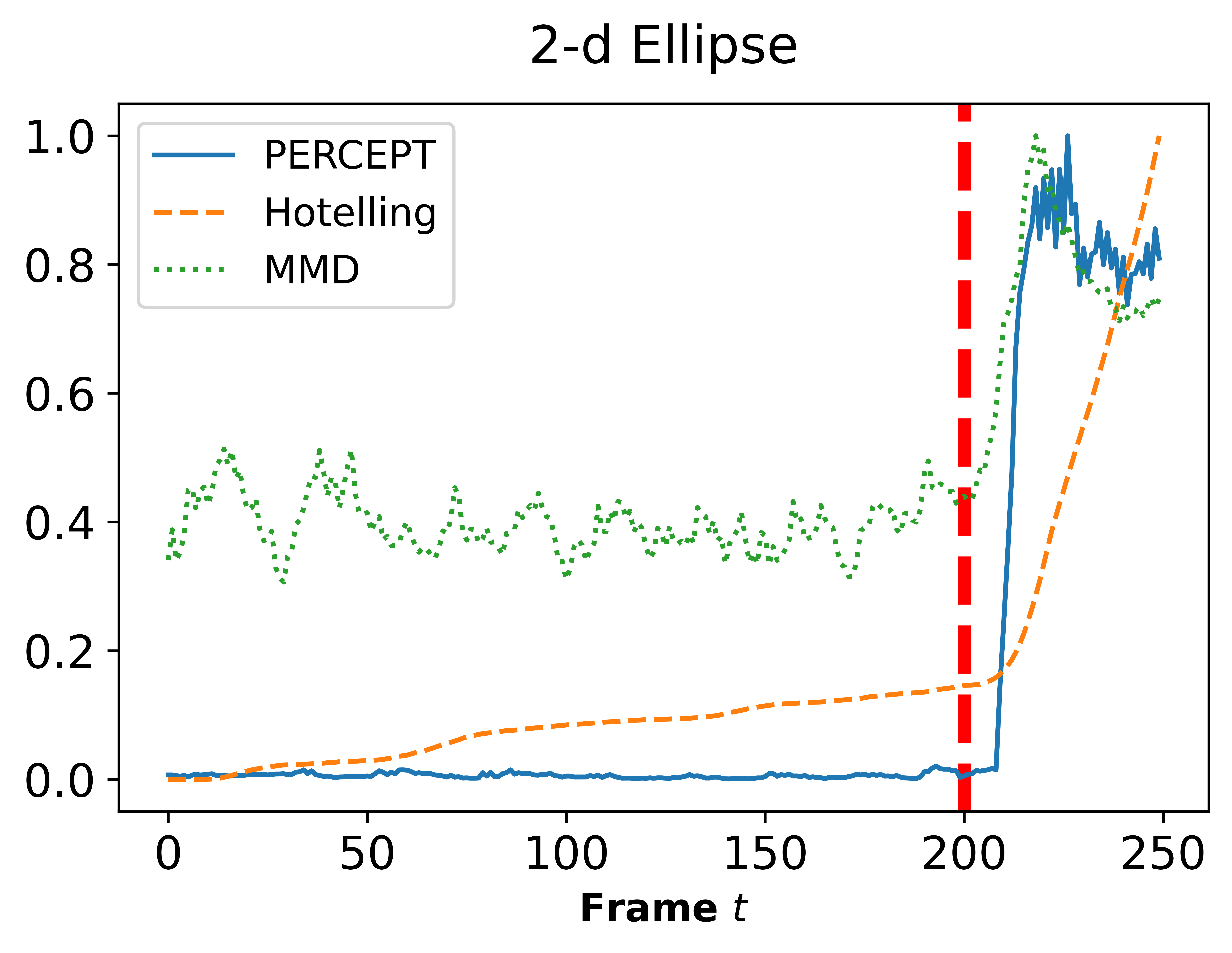} 
    \includegraphics[width=0.45\textwidth]{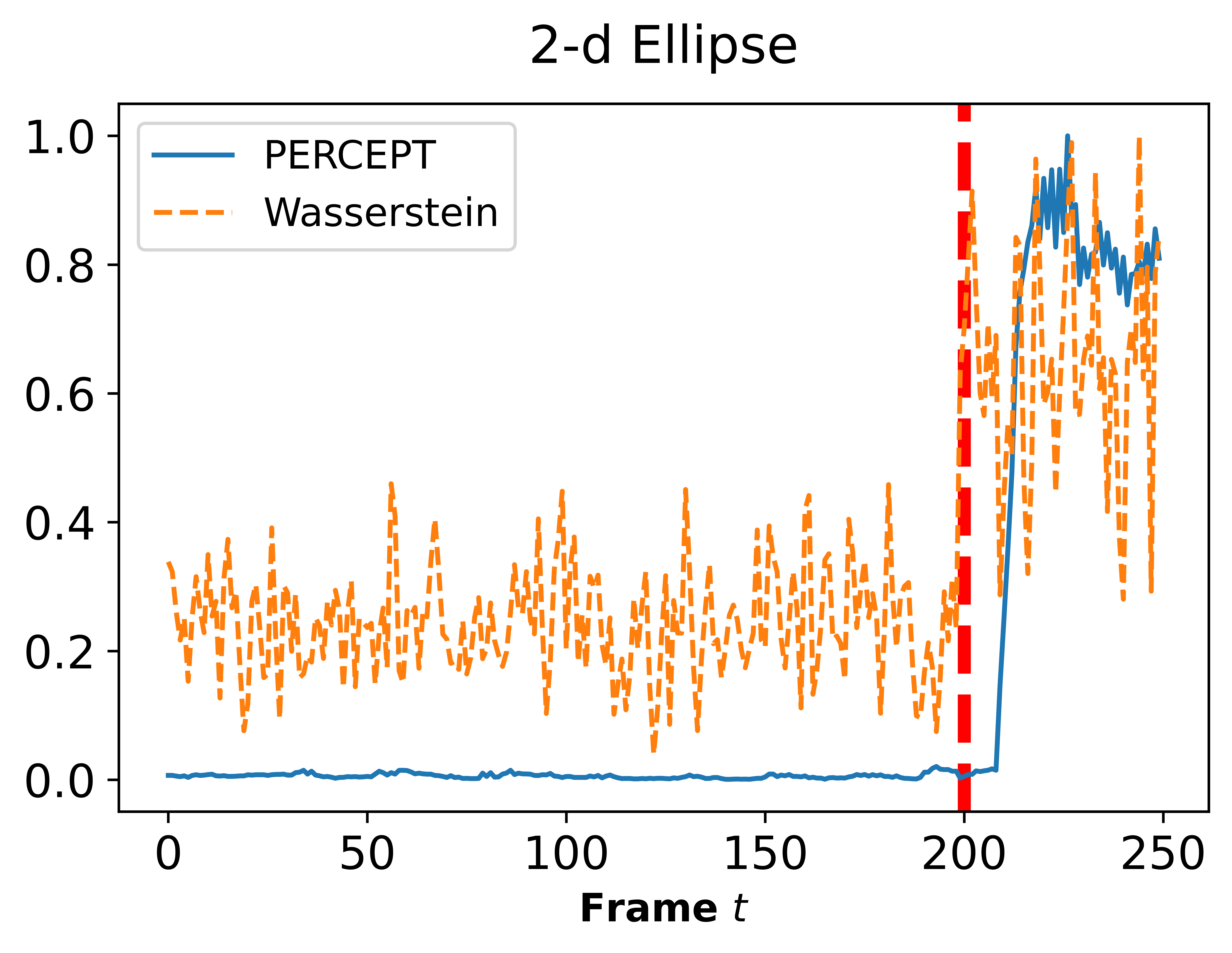}\\
    (b)
    \end{tabular}
    \caption{(a) The $\ell_2$ detection statistic $\chi_t^{max}$, Hotelling's $T^2$ statistics, MMD statistics and the Wasserstein distance at each time $t$ for the 2-D circle with the vertical red dashed line indicating the true change-point, where the noise changes from $N(0, 0.05)$ to $N(0, 0.10)$. (b) Same for 2-D ellipse.}
    \label{noise-change}
\end{figure}

Next, we consider the case of noise changes, where the pre- and post-change data are generated from the same unit 2-D circle (or ellipse), but with noise levels $N(0, 0.05)$ and $N(0, 0.10)$, respectively. Figure \ref{noise-change} 
shows the test statistics from PERCEPT and the Hotelling's $T^2$ for the circle and ellipse. We see that PERCEPT can quickly detect the underlying change: its pre-change test statistic is quite stable, and it peaks up quickly after the change-point at $t=200$. The Wasserstein distance approach also seems to perform quite well here, although its pre-change statistic is noticeably more unstable, which may lead to increased false alarms (i.e., lower ARL). On the other hand, the Hotelling's $T^2$ statistic increases noticeably more slowly after the change, which suggests a larger detection delay. For the MMD statistic, we see that while it peaks up after the change-point, its pre-change statistic is quite unstable and volatile, which again leads to increased false alarms.


\vspace{-0.22in}
\subsection{Increasing Dimensionality}
\vspace{-0.06in}

\begin{figure}[!t]
\vspace{-0.4in}
    \centering
    \begin{tabular}{cc}
    \includegraphics[width=0.32\textwidth]{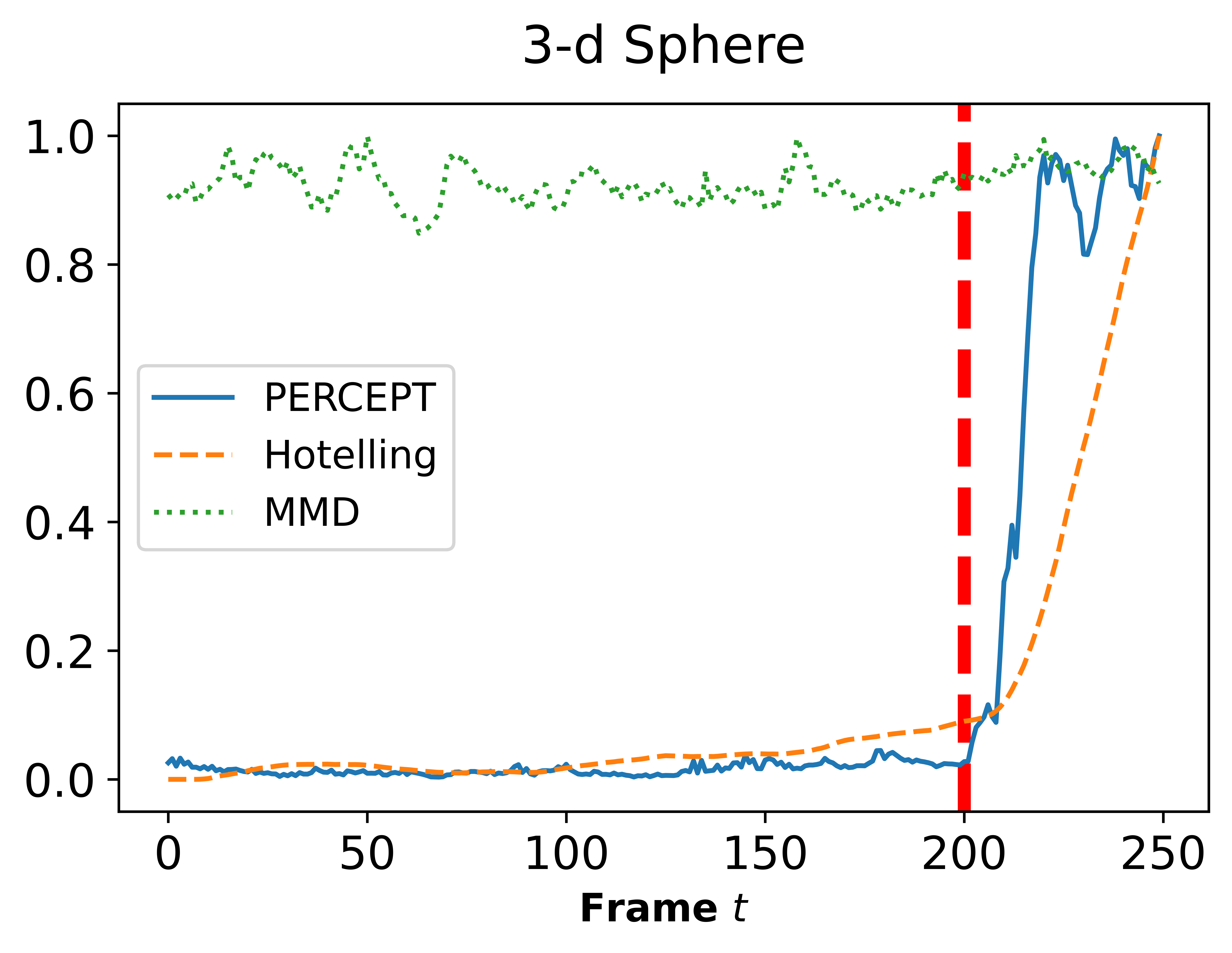}
    \includegraphics[width=0.32\textwidth]{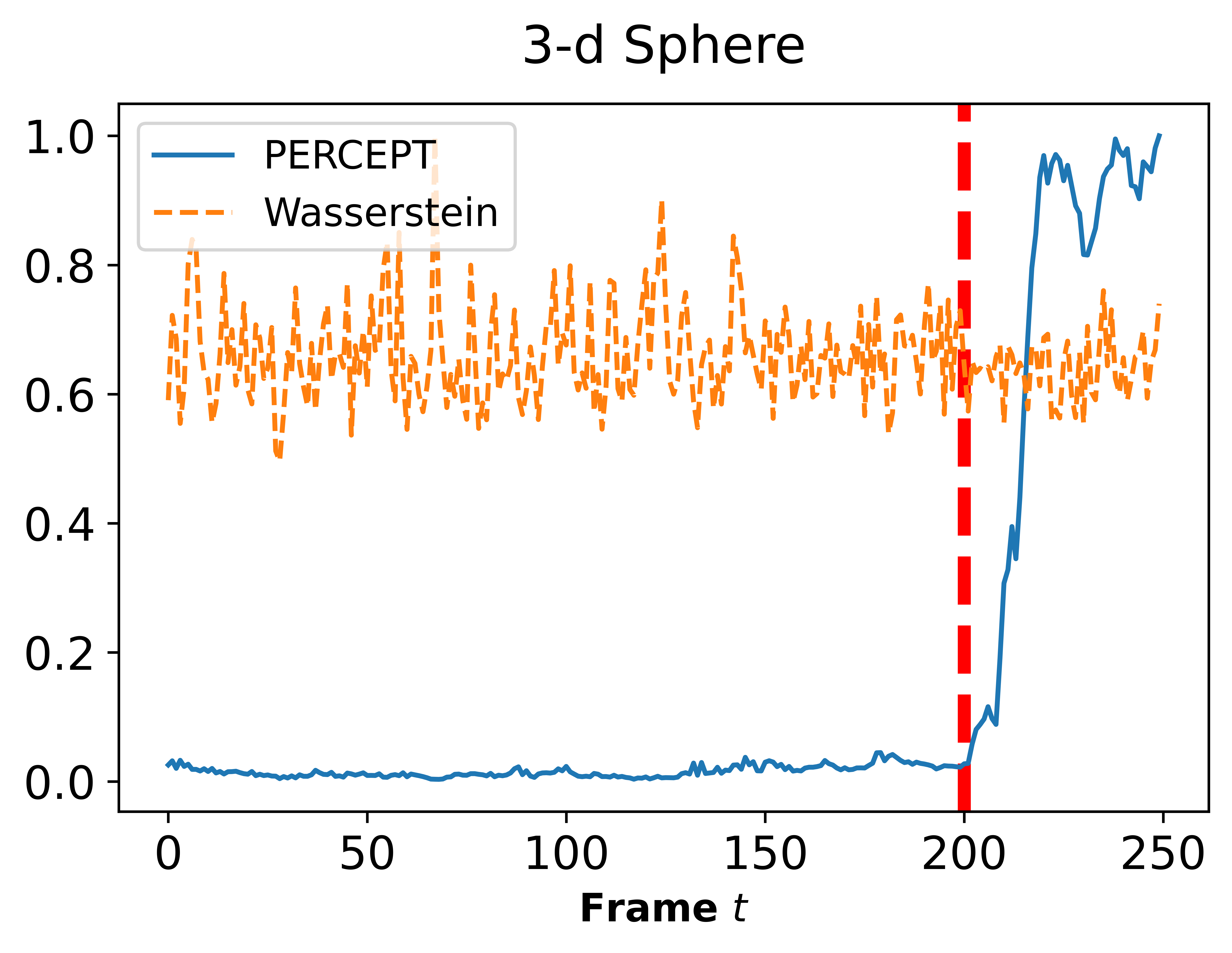}\\
    (a)
    \end{tabular}
    
    \begin{tabular}{cc}
    \includegraphics[width=0.32\textwidth]{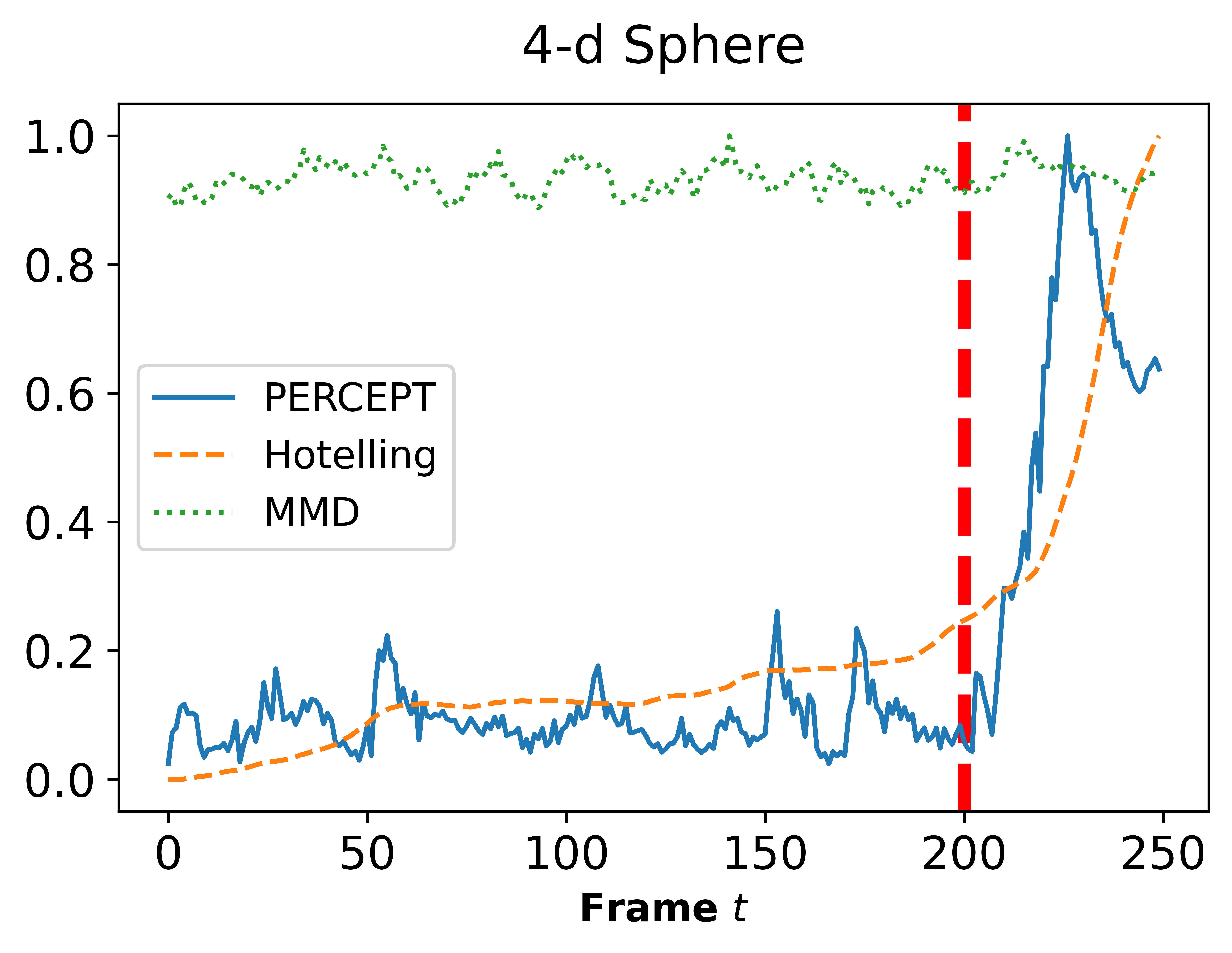} 
    \includegraphics[width=0.32\textwidth]{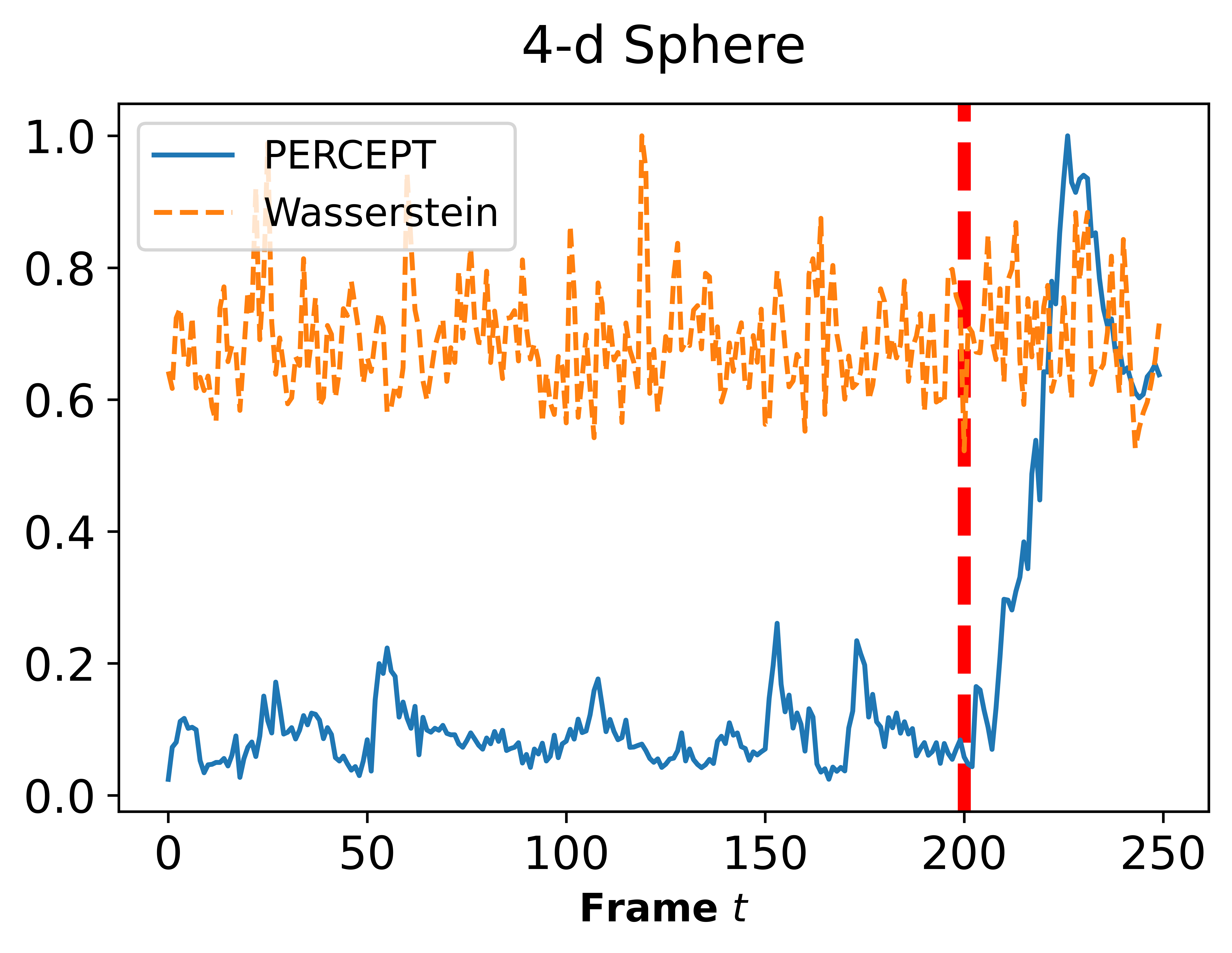}\\
    (b)
    \end{tabular}
    
    \begin{tabular}{cc}
    \includegraphics[width=0.32\textwidth]{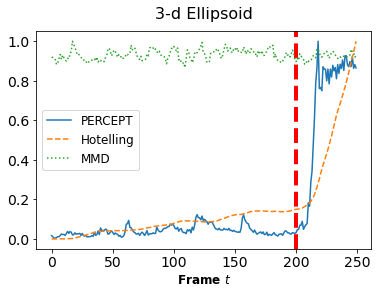}
    \includegraphics[width=0.32\textwidth]{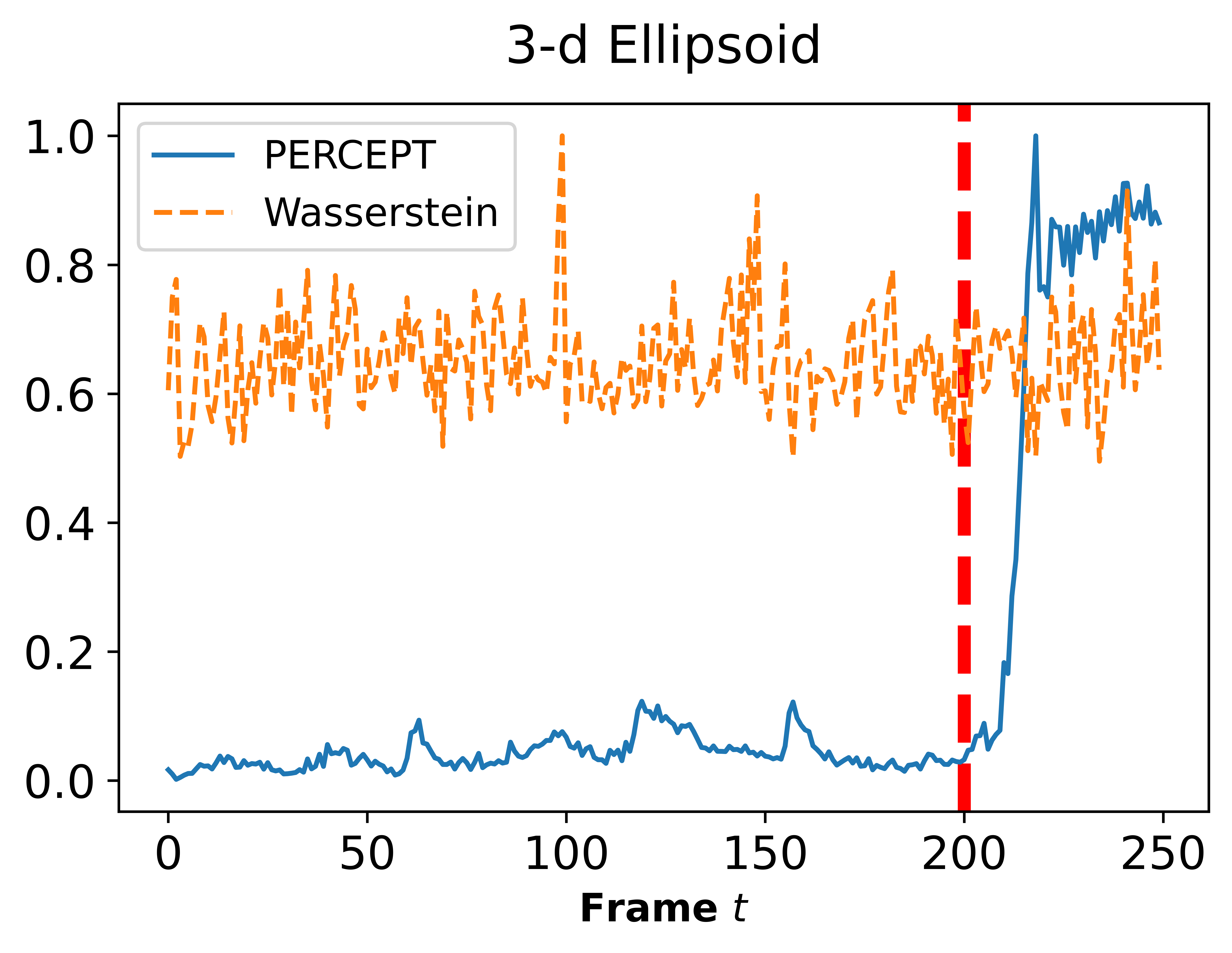}\\
    (c)
    \end{tabular}
    
    \begin{tabular}{cc}
    \includegraphics[width=0.32\textwidth]{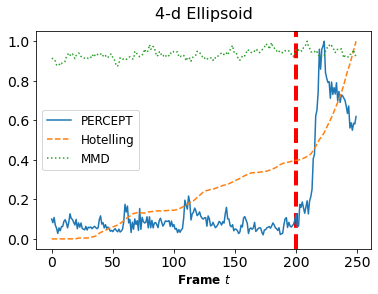} 
    \includegraphics[width=0.334\textwidth]{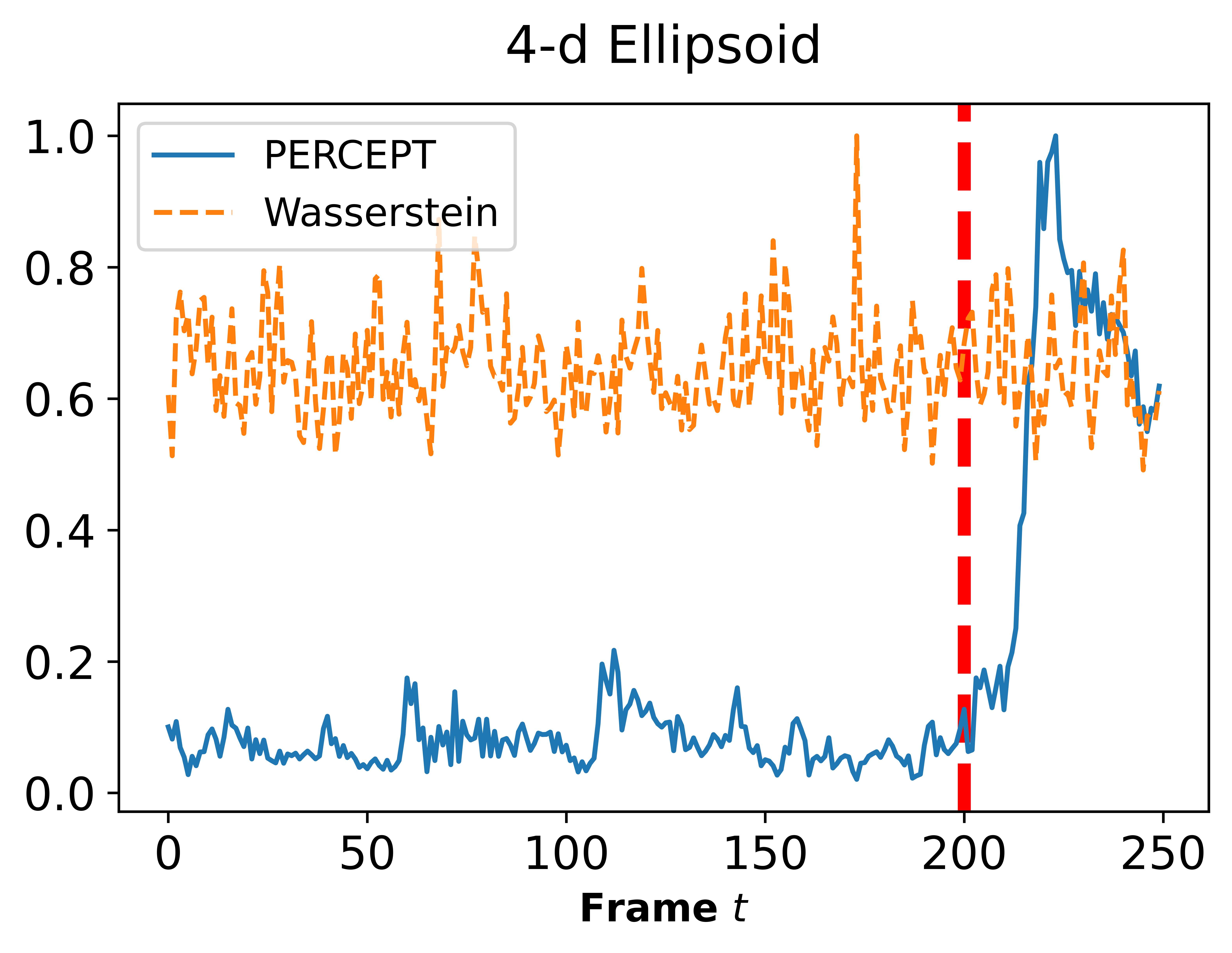}\\
    (d)
    \end{tabular}

    \caption{(a) The $\ell_2$ detection statistic $\chi_t^{max}$, Hotelling's $T^2$ statistics, MMD statistics and the Wasserstein distance at each time $t$ for the 3-D sphere with the vertical red dashed line indicating the true change-point, where the noise changes from $N(0, 0.05)$ to $N(0, 0.10)$. (b) Same for 4-D sphere. (c). Same for 3-D ellipsoids. (d). Same for 4-D ellipsoids.}
    \label{high_dim_sphere}
\end{figure}

We now investigate how well these methods perform on data generated on higher-dimensional geometric structures, namely, the 3-D and 4-D unit spheres and ellipsoids. We consider here the same noise change as in Section \ref{sec:numerical_noise}, and we compared the performance of PERCEPT with the classic parametric Hotelling's $T^2$ test, MMD statistics and Wasserstein distance. Figure \ref{high_dim_sphere}a-b shows the detection statistics for the 3-D and 4-D spheres. We see that the proposed PERCEPT method consistently outperforms existing methods: its pre-change statistics are stable, and its post-change statistics peak up quickly after the change. Comparatively, the increase in the Hotelling-$T^2$ statistic is more subdued after the change (which results in greater detection delay), and the MMD statistic pre-change is noticeably more unstable (which results in increased false alarms). The Wasserstein distance approach is not able to detect the change. Furthermore, comparing with the 2-D results in Figure \ref{noise-change}, the proposed method appears to yield improved performance to existing methods, which is unsurprising since it leverages the underlying low-dimensional topological structure in the high-dimensional data. Figures \ref{high_dim_sphere}c-d show the detection statistics for the 3-D and 4-D ellipsoid. Again, we see that PERCEPT yields improved performance over the existing benchmarks, despite having slightly more unstable pre-change statistics. These experiments suggest that, by learning and integrating low-dimensional topological structure, the proposed method can efficiently detect changes in high-dimensional data.

\vspace{-0.22in}
\subsection{ARL vs. EDD}
\vspace{-0.06in}

Finally, we investigate the performance of these methods via the two metrics introduced in Section \ref{sec:theory}: the Average Run Length (ARL) and the Expected Detection Delay (EDD). Recall that the ARL measures the expected run length to a false alarm when there is no change, the EDD measures the expected number of samples needed to detect a change. We approximate the ARL and EDD of the compared methods using different thresholds $b$, under the 2-D circle experiment with Gaussian noise $N(0,0.05)$ for pre-change regime and Gaussian noise $N(0,\sigma^2)$ for post-change regime, where $\sigma^2 = 0.09, 0.10, 0.11$. We adopted the experiments from \cite{yao_cp}, and details can be found in Appendix \ref{sec:ARLEDD}. 
 

Figure \ref{dim_result} plots the log-ARL vs. EDD comparison for the compared methods in the 2-D circle experiment. Here, a method with large ARL and small EDD is desired, since this results in less false alarms and smaller detection delay. There are two interesting observations to note. First, for all noise levels and any fixed ARL level, we see that PERCEPT yields much lower EDD compared to the Hotelling's $T^2$ procedure, which suggests that the proposed method indeed yields improved detection performance by integrating topological structure. Second, the proposed method appears to be much more robust to noise perturbations. As the noise level decreases, the EDD for the Hotelling's $T^2$ becomes noticeably larger for fixed ARL levels. This is because the noise change becomes smaller as we decrease the noise level of the post-change data. This is not surprising, since this change is difficult to detect without first identifying the underlying topological structure. In contrast, the EDD for PERCEPT is more stable and nearly remains the same as the noise level increases, which shows the robustness of PERCEPT. The relationship between ARL and EDD for the 4-D case is not shown here because it's computational expensive, but the same conclusion is expected from the earlier results.


\begin{figure}[!t]
\vspace{-0.2in}
    \centering
    \begin{tabular}{cc}
     \includegraphics[width=0.455\textwidth]{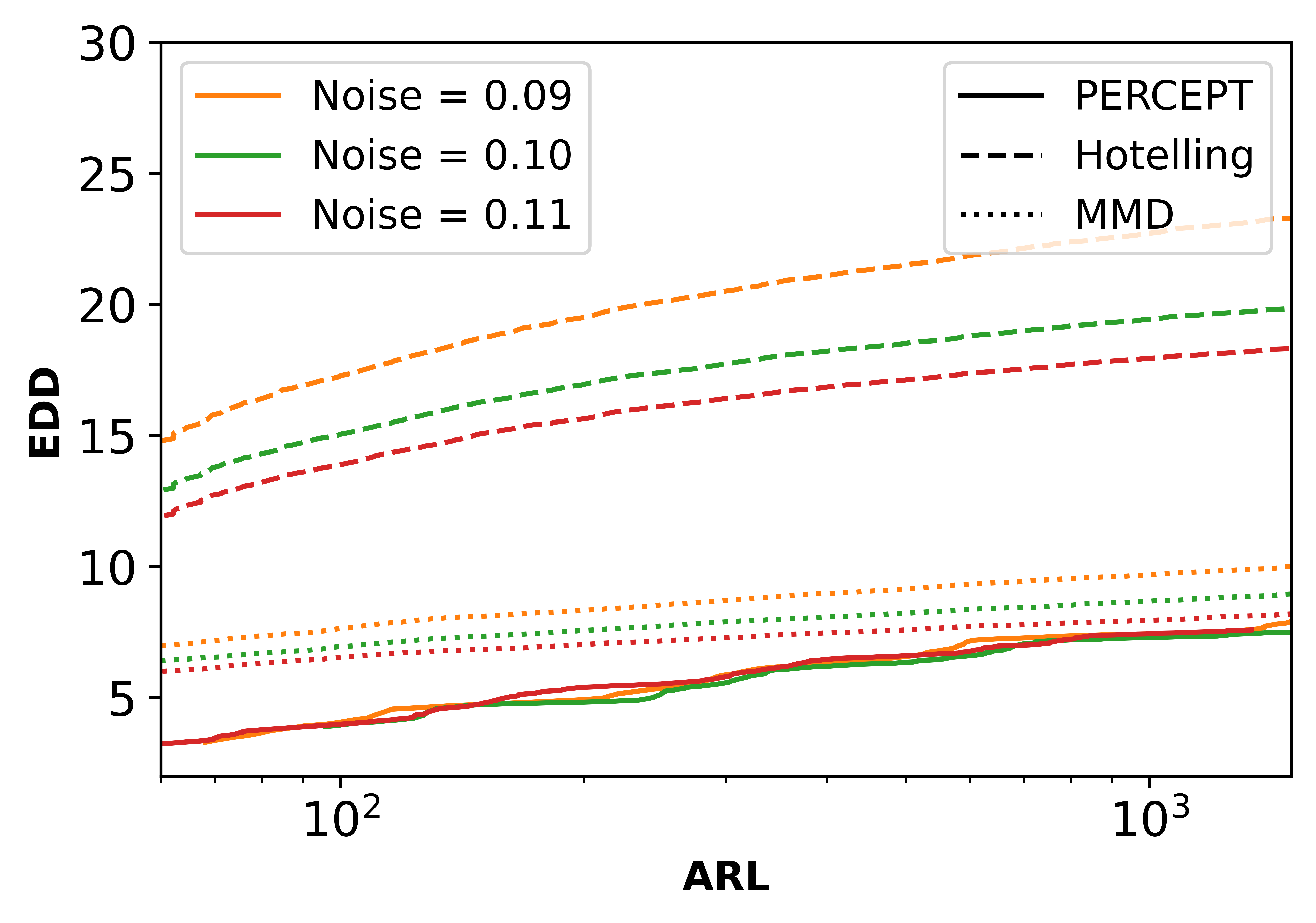}    & 
     \includegraphics[width=0.465\textwidth]{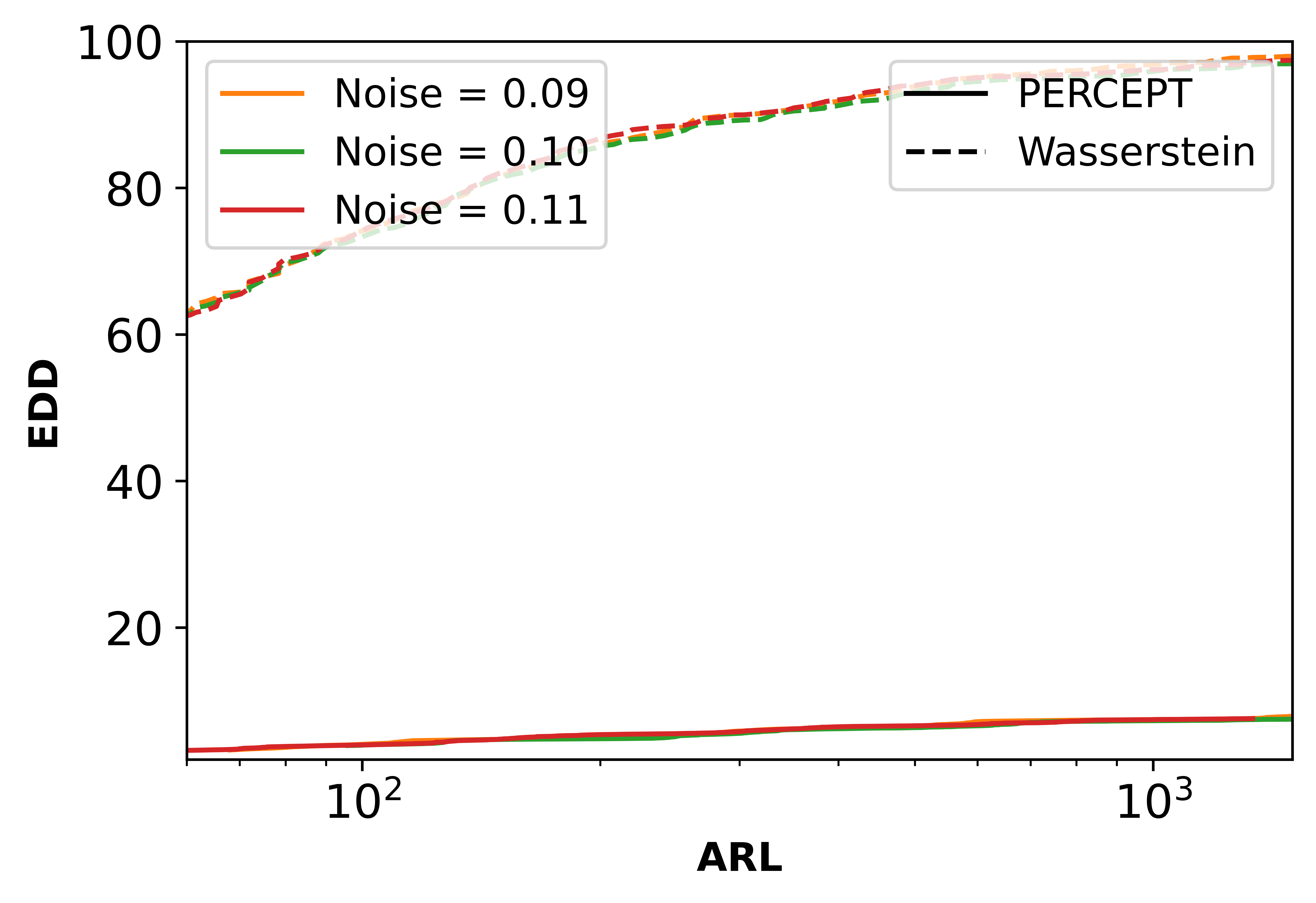}  \\
    \end{tabular}
    \caption{EDD--ARL curves for 2-D circle with varying noise levels for PERCEPT, the Hotelling's $T^2$, the MMD test and the Wasserstein distance procedure, respectively.}
    \label{dim_result}
\end{figure}

\vspace{-0.2in}
\section{Applications}\label{sec:data}

\vspace{-0.22in}
\subsection{Solar Flare Change Detection}
\vspace{-0.06in}

We now demonstrate the effectiveness of PERCEPT in the earlier motivating problem on detecting changes in solar flares. A solar flare is an intense emission of radiation on the Sun's atmosphere, and the monitoring of such changes via satellite imaging is critical for predicting geomagnetic storms \citep{yao_solar}. This detection can be highly challenging, however, as noted in Section \ref{sec:exist}, due to the high dimensionality of image data and the subtlety of such a change. Recent work in image processing suggests a wide range of image features can be captured via topology \citep{image_ls}, which suggests our topology-aware approach may provide a potential solution. The data used here are $T=100$ image snapshots taken by the SDO at NASA, where the true change-point is at $t^* = 49$. Further details can be found in Section \ref{sec:exist}.

\begin{figure}[!t]
\vspace{-0.3in}
     \centering
    \includegraphics[width=0.9\textwidth]{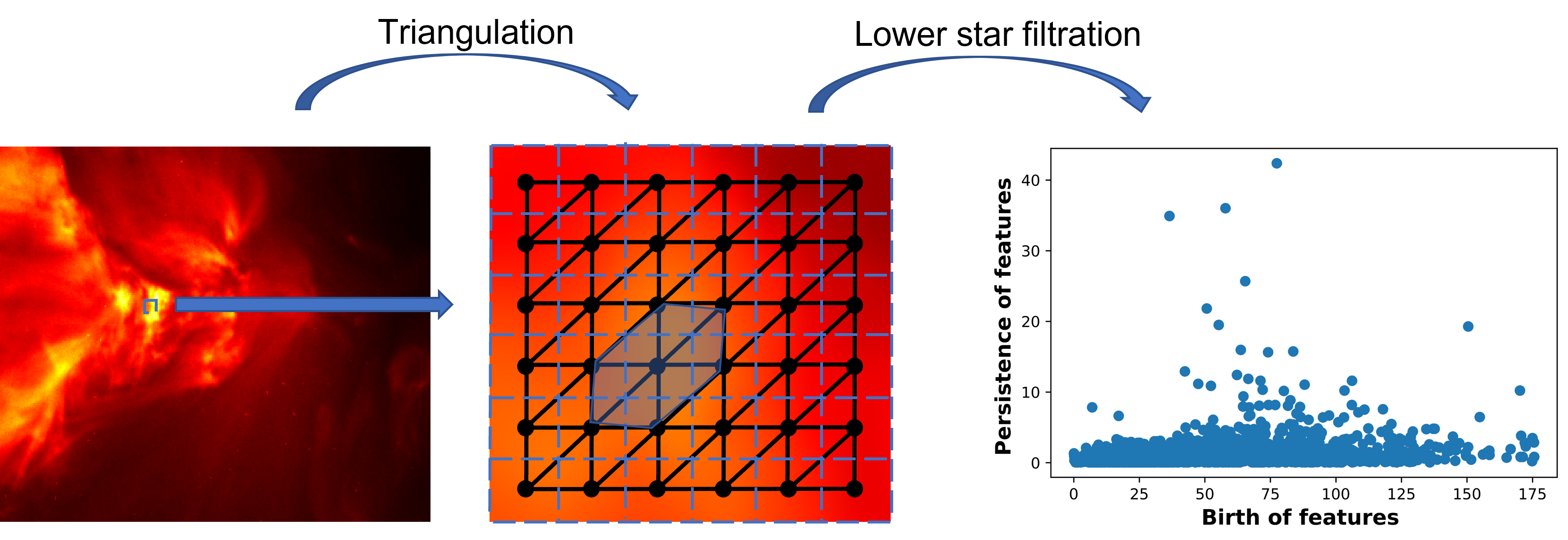} \\
    \hspace{0.6in}
    (a)
    \hspace{3.2in}
    (b)
     \caption{(a) Triangulation of an image. (b) ``Tilted" persistence diagram of the solar flare image in (a).
     }
        \label{fig solar graphs}
\end{figure}

To apply PERCEPT, we need to first map the image data to a filtration, or a sequence of simplicial complexes (see \ref{sec:tda} for details). We will make use of the so-called \textit{lower star filtration}, which has shown success in capturing useful topological features in image data \citep{image_ls}. Let $f(x)$ be a mapping from each pixel location $x$ to its intensity value, and for a given $\epsilon$, define the \textit{sublevel set} of $f$ as $X(\epsilon) = \{x | f(x) \leq \epsilon\}$. One can then form a simplicial complex from $X(\epsilon)$ by first considering image pixels as vertices on a grid, then triangulating this grid by placing an edge between two points that are horizontally, vertically, or diagonally adjacent, and a triangular face for any three adjacent points forming a triangle. Figure \ref{fig solar graphs}a visualizes this mapping from image to simplicial complex. For a sequence $ 0< \epsilon_1 <\epsilon_2< \cdots <\epsilon_m$, the \textit{sublevel set filtration} of the image is then defined as $X_i = X(\epsilon_i)$, $i = 1,\cdots,m$, which forms a sequence of nested simplicial complexes. This filtration can thus be used to construct a persistence diagram (Figure \ref{fig solar graphs}b) as described in Section \ref{sec:tda}. One can view this filtration pipeline as a way to extract important image topological features as a PD point cloud.



With the mapped PD in hand for each image, we can then proceed with the detection framework outlined in Section \ref{sec:PERCEPT}. A quick inspection of the computed PDs show only a small amount of points (or features), thus we decided to use persistence histograms instead of persistence clusters (see Section \ref{sec:voronoi}). Here, the histogram breakpoints $b_1, \cdots, b_L$ are chosen such that there is (roughly) an equal sum of persistences within each histogram bin in the first solar flare image. The Hotelling's $T^2$ is performed using the $30$ extracted principal components from PCA on the image data, and the MMD test is performed on the image data directly, with the RBF kernel, and the bandwidth is chosen using the ``median trick" as describe in Section \ref{sec:num}. The Wasserstein distance approach is performed on the mapped PDs from lower star filtration. 

\begin{figure}[!t]
\vspace{-0.2in}
    \centering
    \begin{tabular}{cc}
     \includegraphics[width=0.45\textwidth]{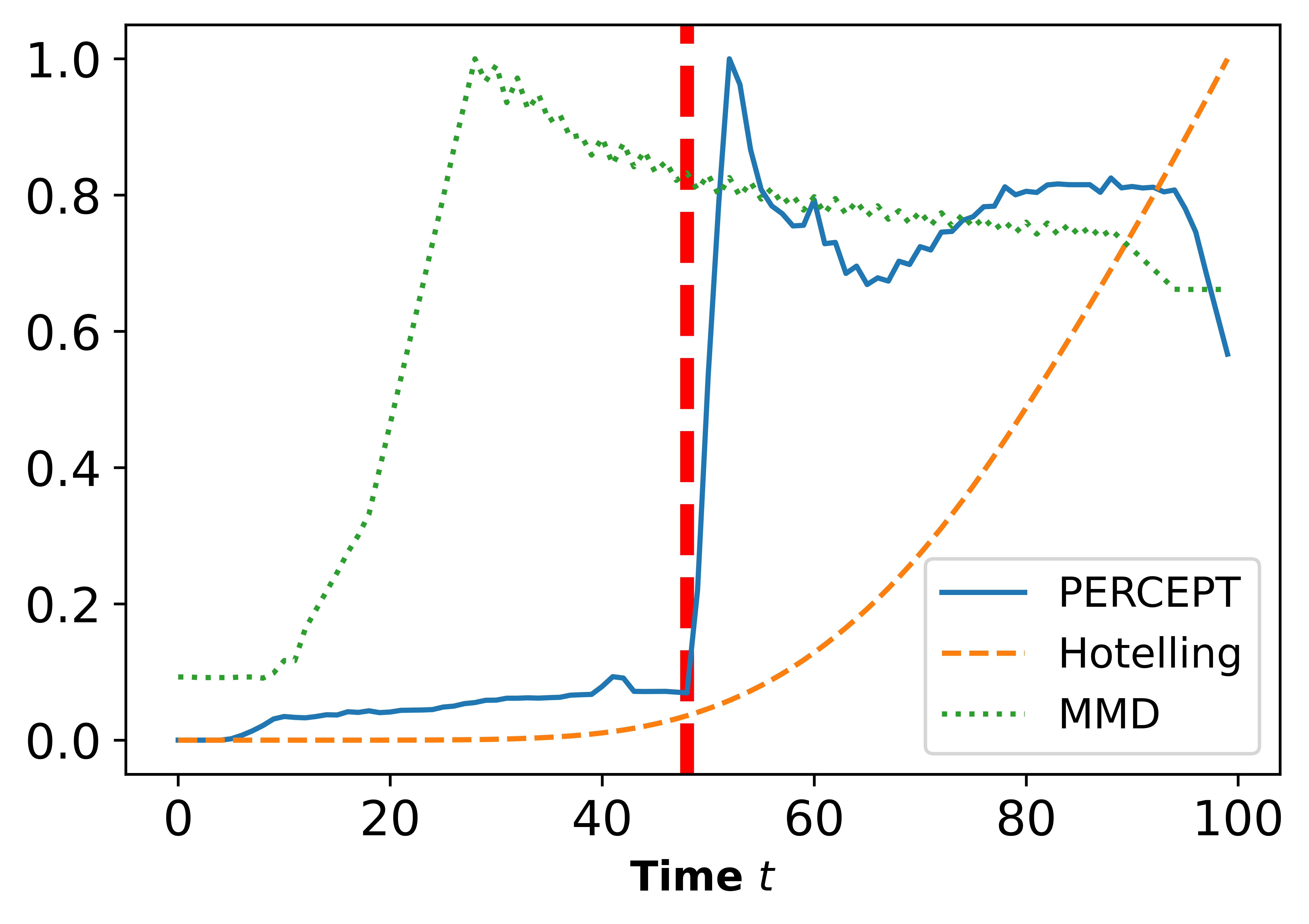}
     \includegraphics[width=0.45\textwidth]{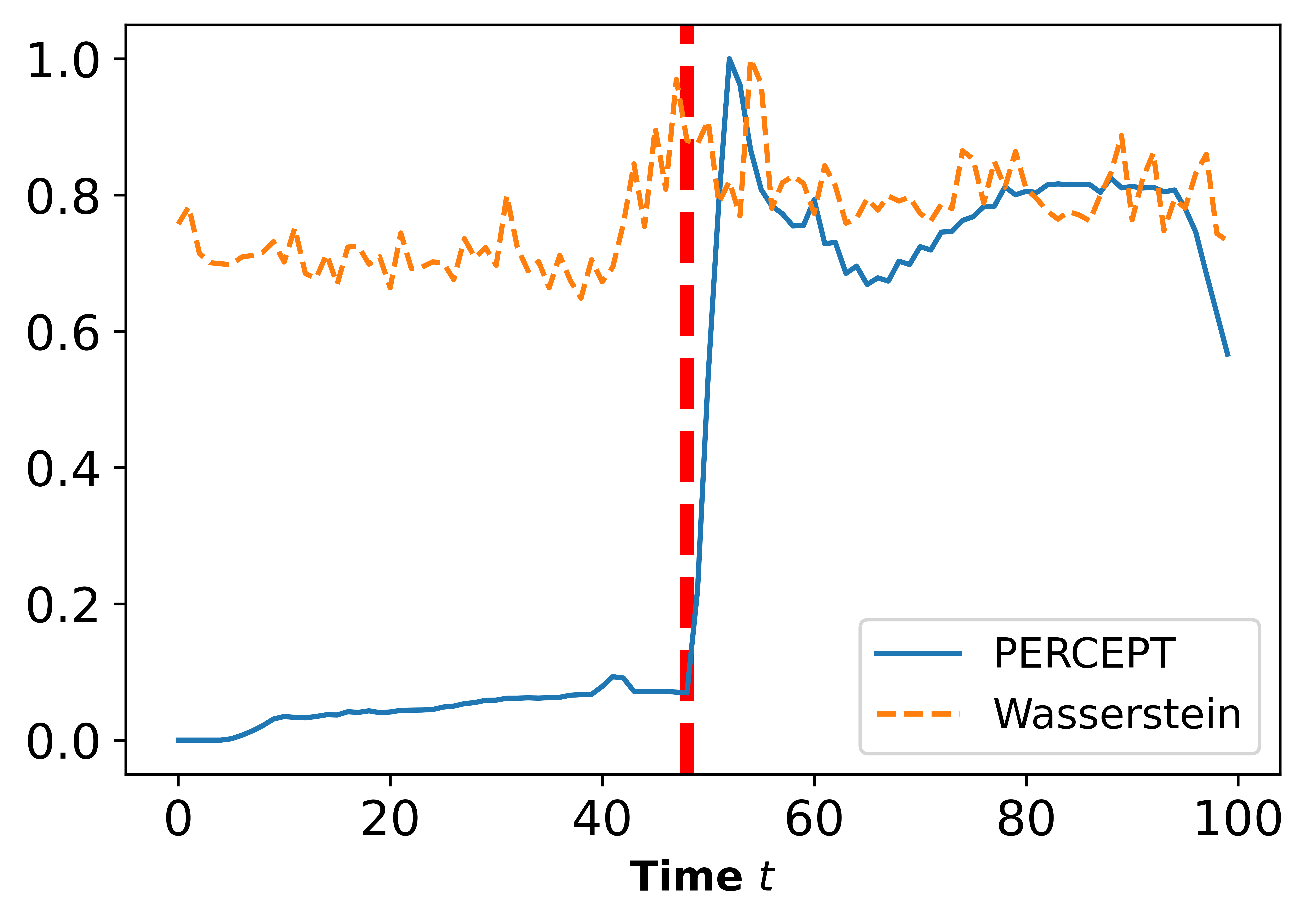}
    \end{tabular}
    \caption{Detection statistics for each compared method in the solar flare monitoring application. The vertical red dashed line indicates the true change-point.}
    \label{fig:three graphs}
\end{figure}

Figure \ref{fig:three graphs} shows the proposed detection statistic $\chi_t^{\max}$ using $L=10$ histogram bins, the Hotelling's $T^2$, the MMD test statistics and the Wasserstein distance over time. For PERCEPT, we note a sudden increase in the test statistic after the true change-point $t^*=49$ (vertical red dashed line), which suggests the proposed method is able to pick out the underlying topological change in the images with little detection delay. Comparatively, for the Hotelling's $T^2$, the increase in its test statistic is much more subdued and gradual, which indicates a much larger detection delay. The MMD approach yields poor performance here: its pre-change statistics are highly volatile and unstable, and its post-change statistics experience a decrease after the change-point. The Wasserstein distance approach is not able to detect the change: there is no large change in statistics at the change-point. This shows that, by learning and integrating low-dimensional topological structure within a non-parametric change detection framework, the proposed approach can yield significant improvements over existing methods when such structure is indeed present for image data.


We also note that, despite the required mapping from images to PDs via the lower star filtration, PERCEPT is quite computationally efficient. This is due in large part to the availability of well-maintained packages for TDA algorithms. Here, using the Python package \texttt{Ripser} \citep{ripser}, the computation time for performing the lower star filtration on $300$ images is approximately $90$ seconds on a standard desktop computer. Given this computed filtration, the detection statistic $\chi_t^{\max}$ can then be evaluated with minimal additional computation, thus allowing for efficient topology-aware online detection.

\vspace{-0.22in}
\subsection{Human Gesture Change Detection}
\vspace{-0.06in}

\begin{figure}[!t]
\vspace{-.4in}
    \centering
    \includegraphics[width = 0.9\textwidth]{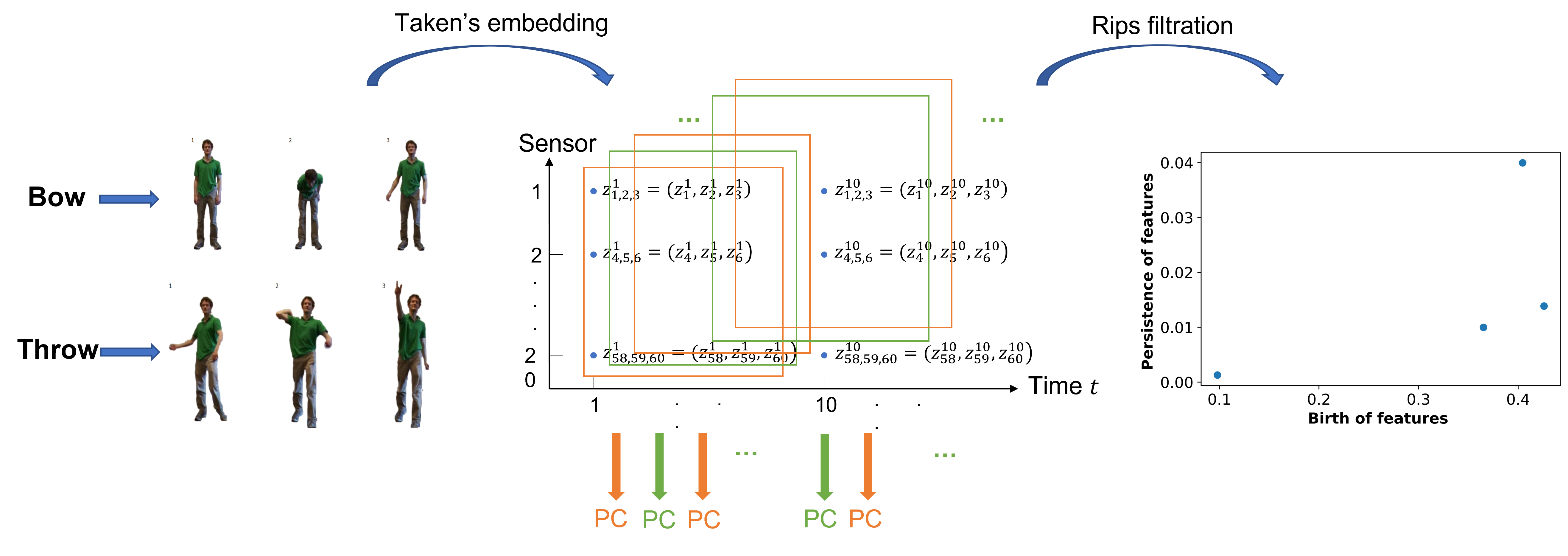}\\
    \hspace{0.3in}
    (a)
    \hspace{1.7in}
    (b)
    \hspace{2.15in}
    (c)
    \caption{(a) Visualizing the two gestures: taking a bow (top) and throwing an object (bottom). (b) Embedding of a multivariate time series to point cloud (PC) data to a persistence diagram. (c) ``Tilted" persistence Diagram of the embedded point cloud at $t=510$. }
    \label{gesture_stat}
\end{figure}

Next, we investigate the performance of PERCEPT in a human gesture detection application. The detection of human body gestures is an important task in computer vision \citep{comp_vision}, and has immediate applications in visual surveillance and sign language interpretation \citep{visual_surveillance}. One challenge for this detection lies in the high-dimensional time series data, and the low-dimensional embedding of human body gestures within such data. Recent developments in time series analysis suggest that many time series features can be captured via TDA (further details below), and thus PERCEPT may be promising for this task. To explore this, we will use the human gestures dataset from the Microsoft Research Cambridge-12 Kinect \citep{gesture_data}, which consists of time series observations of human skeletal body part movements, collected from 20 sensors on 30 people performing 12 distinct gestures. We study in particular the transition of gestures from a ``bow" to a ``throw" sequence, as shown in Figure \ref{gesture_stat}a.

To apply PERCEPT, we need to first transform the multi-dimensional time series to point cloud data, on which the usual TDA pipeline (see Section \ref{sec:tda}) can then be performed. A popular transformation for this is Taken's embedding \citep{ts_embedding}, which is widely used in TDA. Suppose we observe the $d$-dimensional time series $\{z_k(t), t \geq 0\}$, $k = 1, \cdots, d$ (here, $d=20 \times 3 = 60$, since each of the 20 body sensors return a 3-D coordinate). For each time $t$, define the point $\bm{z}_t = (z_1(t), \cdots, z_d(t))$ as the cross-section of the multivariate time series. With a given window size $w$, Taken's embedding returns the point cloud representation $\mathcal{Z}_t = \{\bm{z}_{i}, \cdots, \bm{z}_{i+w-1}\}$ at each time $t$. After this transformation from a $d$-dimensional time series to the point cloud $\mathcal{Z}_t$, one can then perform the standard TDA filtration in Section \ref{sec:tda} to convert $\mathcal{Z}_t$ to a persistence diagram $\mathcal{D}_t$. It can be shown \citep{ts_embedding} that this embedding captures key periodicity and dynamic system information on the multivariate time series. Figure \ref{gesture_stat}b provides a visualization of Taken's embedding. Here, we set the window size $w$ to be approximately the periodicity of the gestures, which is based on prior knowledge.


With this, we can then apply PERCEPT for detecting gesture changes. A quick inspection shows that there are limited points in the computed PDs $\mathcal{D}_t$. Thus we use the persistence histograms in Section \ref{sec:perhist} rather than the persistence clusters in Section \ref{sec:voronoi}. We then used 60 frames from ``bow" and ``throw" sequences to choose the number of histogram bins and optimize weights. The Hotelling's $T^2$ is performed using the $30$ extracted principal components from PCA on the embedded point cloud data $\mathcal{Z}_t$, and the MMD test is performed on the embedded point cloud data $\mathcal{Z}_t$ directly, with the RBF kernel, and the bandwidth is chosen using the ``median trick" as described in Section \ref{sec:num}. The Wasserstein distance approach is performed on the mapped PDs from the Rips filtration. 

\begin{figure}[!t]
\vspace{-0.2in}
    \centering
    \begin{tabular}{cc}
     \includegraphics[width=0.45\textwidth]{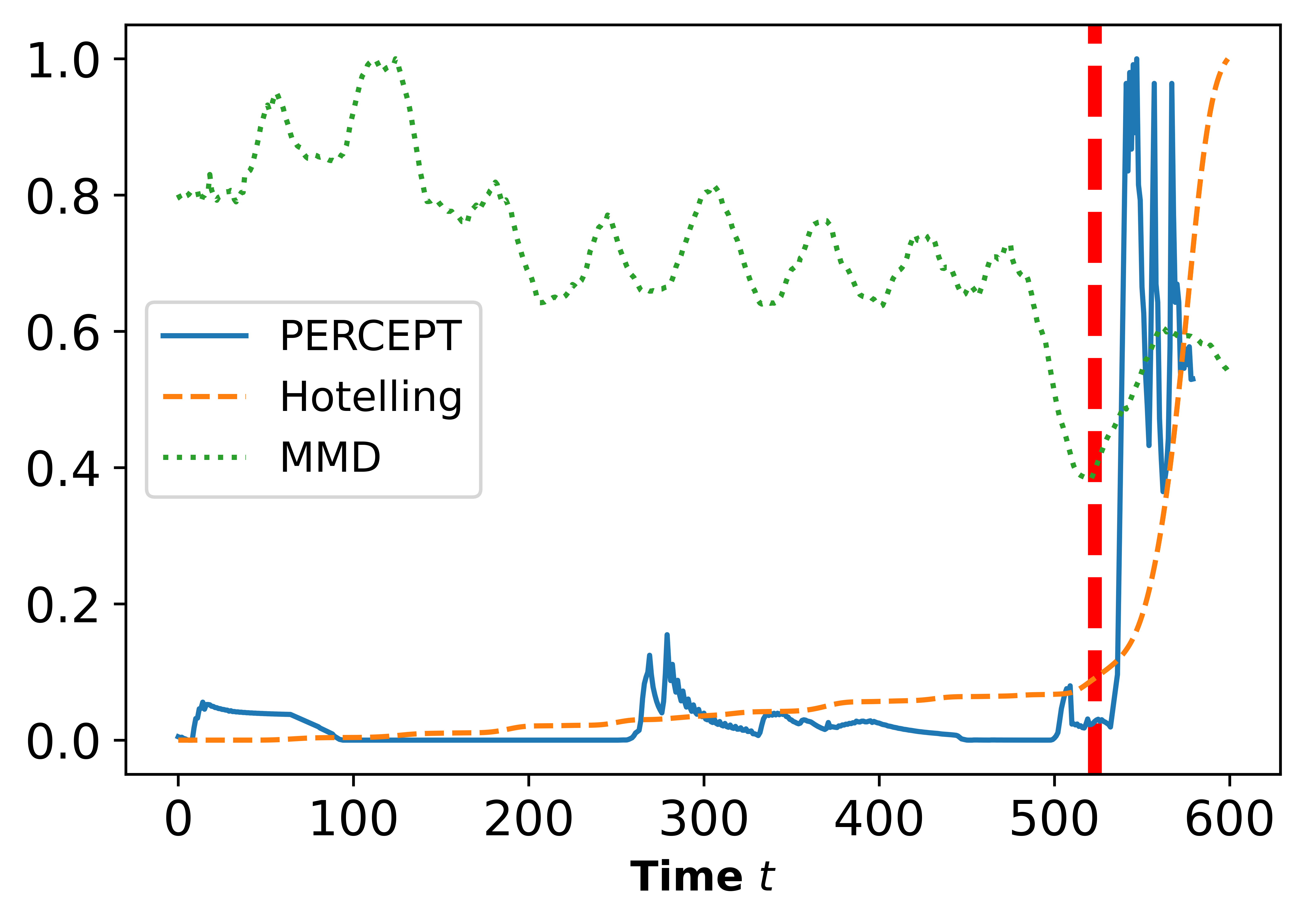}
     \includegraphics[width=0.45\textwidth]{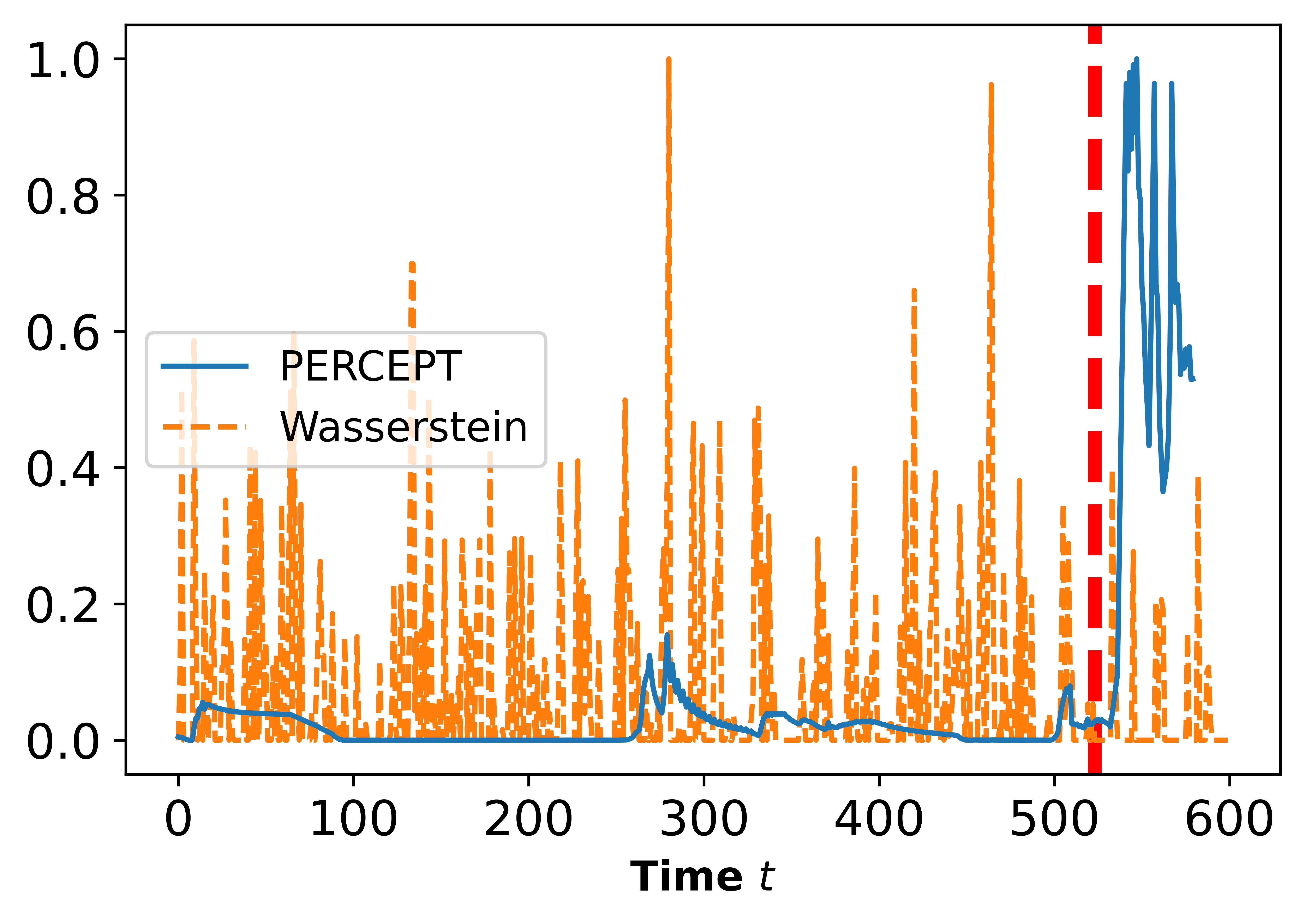}
    \end{tabular}
    \caption{Detection statistics for each compared method in the human gesture change detection application. The vertical dashed line indicates the true change-point from ``bow'' to ``throw''.}
    \label{human_stat}
\end{figure}

Figure \ref{human_stat} shows the proposed detection statistic $\chi_t^{\max}$ using $L=2$ histogram bins, the Hotelling's $T^2$, the MMD test statistics, and the Wasserstein distance over time, with the vertical line indicating the true change-point (i.e., the transition from ``bow'' to ``throw''). For PERCEPT, we see the test statistic is relatively stable pre-change and experiences a large jump immediately after the change. This suggests our method has a large ARL and small detection delay, which is as desired. The Hotelling's $T^2$ again experiences a much more subtle and gradual increase after the change, which indicates a large detection delay. As before, the MMD method performs poorly: its pre-change statistic is highly unstable pre-change and experiences a notable decrease post-change, which suggests it is unable to detect the change at all. The Wasserstein distance is again not able to detect the change-point. The distance hits zero regularly, indicating the similar topological structure between the adjacent embedded point clouds, which shows the importance of using historical data for change-point detection. This demonstrates the advantages of a topology-aware non-parametric change detection framework: when such a low-dimensional structure exists and can be leveraged, one can achieve efficient monitoring performance even with high-dimensional data.


\vspace{-0.2in}
\section{Conclusion}
\vspace{-0.06in}

We have proposed a novel topology-aware, non-parametric monitoring framework called the Persistence Diagram-based Change-Point Detection (PERCEPT) method, which leverages tools from topological data analysis for change detection. The idea is to first extract the topological structure of the data via persistence diagrams, then leverage a non-parametric, histogram-based change detection approach on these diagrams to sequentially detect topological changes. A suite of simulation experiments and two applications show that, when the underlying change is topological in nature, the proposed PERCEPT approach yields noticeably improved detection performance over existing approaches.

Despite these promising results, there are several interesting future directions for refining PERCEPT for broader applications. First, we are exploring a more localized detection approach, which can pinpoint and monitor local changes (e.g., local translation/rotation shifts) in topology. Second, there has been recent work on two-parameter persistence \citep{Wright_Zheng_2020}, particularly on its robustness in extracting topological structure in the presence of noise; integrating such ideas within PERCEPT would allow for a more robust topology-aware monitoring approach. Finally, we will investigate broader uses of PERCEPT in applications for which TDA has found recent success, particularly neuroscience \citep{wang2020uncertainty,wang2021sequential} and complex physical systems \citep{mak2018efficient}.


\if0\blind
\textbf{Acknowledgements}: The authors also thank Yi Ji and Alessandro Zito for helpful comments on the manuscript. Yao Xie is partially supported by NSF DMS-2134037, CCF-1650913, DMS-1938106, and DMS-1830210.
\fi 


\bigskip
\setlength{\bibsep}{3pt}
{\linespread{1}\selectfont\bibliography{references}}


\newpage
\begin{appendices}

\section{Experiments for approximating the ARL and EDD}
\label{sec:ARLEDD}

Experiment 1: To compute the ARL, we generate $n$ sequences of pre-change samples of length $m$. We create a random pool with $M$ pre-change samples to reduce the computational cost. For each sequence, $m$ samples are drawn randomly from the pool with replacement, and we repeat this process $n$ times. Let $\mathcal{T}$ be the stopping time of the detection procedure, and if there is no change-point in the sequence of length $m$ (all samples are drawn from the pre-change distribution), from the discussion in \cite{yao_cp} we have:
$$P(\mathcal{T} > m) = P(\max_{0 \leq t \leq m} \chi_t < b) \approx \exp\{-m/\lambda\},$$
where $\lambda$ is the estimation of the ARL. For any given threshold $b$, we could get the estimation of the ARL based on the percentage of sequences whose maximum online statistics is below the threshold, among all $n$ sequences. More specifically, we approximate the ARL as $m/(-\ln{\hat{p}})$, where $\hat{p}$ is the percentage.

Experiment 2: To compute the EDD, we generate $n$ sequences of pre-change samples of length $m'$ followed by post-change samples of length $m$. The pre-change samples are only used as historical data, in order to construct the pre-change histograms $\boldsymbol{\omega}_{t,k}^{[1,1]}$, $\boldsymbol{\omega}_{t,k}^{[1,2]}$ as shown in Figure \ref{fig:hist_cp}b. The online test statistics are calculated only from the onset of post-change samples to obtain the detection delay. We consider the same list of threshold $b$ as we used in computing the ARL, and find the average detection delay over $n$ sequences. 

\section{Details on the lower star filtration}

\begin{figure}
     \centering
    \includegraphics[width=0.5\textwidth]{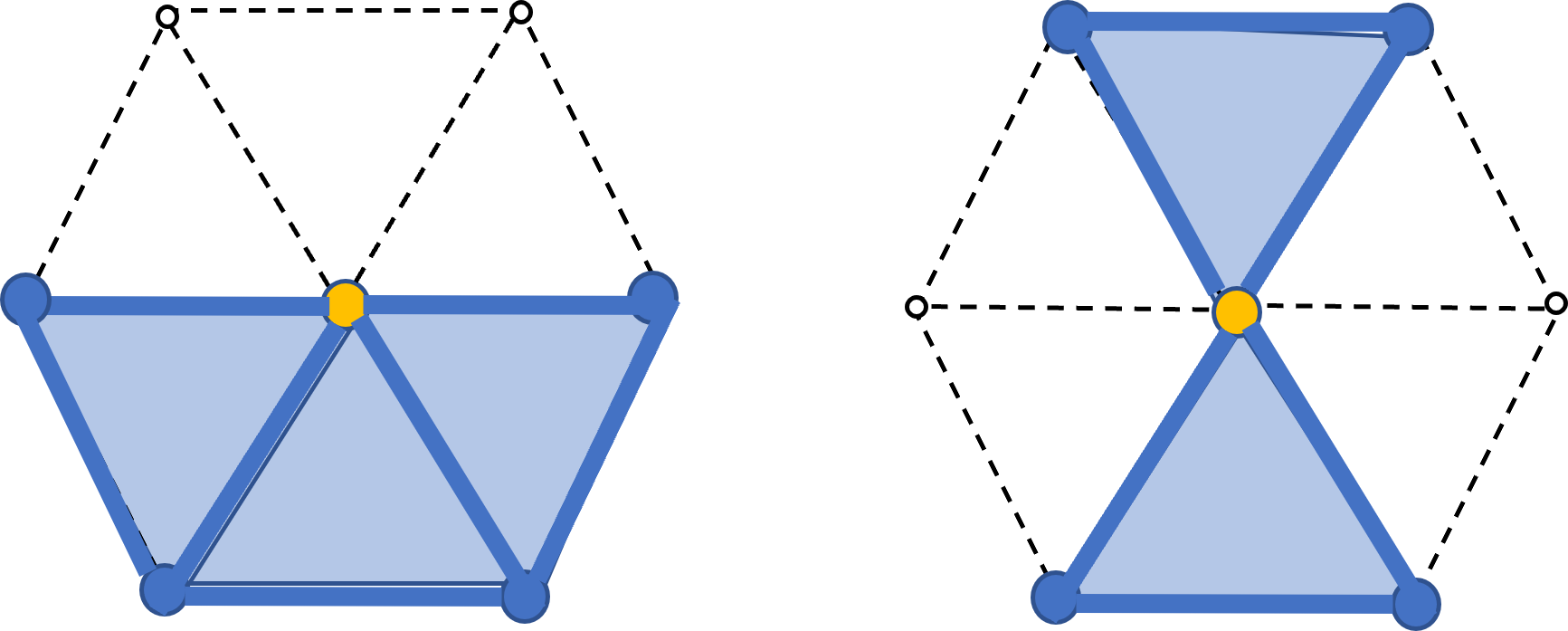}
     \caption{Visualizing a regular point (left) and a saddle point (right) in the lower star filtration.
     }
        \label{appendix2}
\end{figure}

When a new vertex is added in the sublevel set, the topological change depends on whether the vertex is a maximum, minimum, regular, or a saddle of the function. Figure \ref{appendix2} visualizes a regular point and saddle point (in yellow), and the edges and faces in the sublevel sets (in blue). The topological features do not change after introducing a regular point, but the number of connected components decreases by one after introducing a saddle point. 

\section{Connection on distributional \& bottleneck distances}\label{sec:prop}

Here, we elaborate further on the connection between the difference between the pre- and post-change persistence distributions, $\|p_{\rm pre} - p_{\rm post}\|_2$, and its corresponding bottleneck distance $d_B(\mathcal{D}_{\rm pre},\mathcal{D}_{\rm post})$. Since the bottleneck distance is defined on the sampled PDs $\mathcal{D}_{\rm pre}$ and $\mathcal{D}_{\rm post}$ at a given time $t$, we will make this connection on the \textit{empirical} pre- and post-change persistence distributions $\hat{p}_{\rm pre}$ and $\hat{p}_{\rm post}$ at time $t$, respectively.

The following proposition provides a link between the empirical persistence distribution difference $\|\hat{p}_{\rm pre} - \hat{p}_{\rm post}\|_2$ and the bottleneck distance of its corresponding persistence diagrams $\mathcal{D}_{\rm pre}$ and $\mathcal{D}_{\rm post}$, under asymptotic conditions.

\begin{prop}

Let $\mathcal{D}_{\rm pre}$ and $\mathcal{D}_{\rm post}$ be the PDs of samples from pre- and post-change data at time $t$, and let $\hat{p}_{{\rm pre}}$ and $\hat{p}_{{\rm post}}$ be the corresponding persistence histograms given a fixed partition of $L$ bins at time $t$. 
Suppose all birth times are unique, and suppose the number of histogram bins $L$ are sufficiently large such that the persistence histograms at each time has at most one point in each bin. Then $||\hat{p}_{{\rm pre}}-\hat{p}_{{\rm post}}||_2^2 \geq d^2_B(\mathcal{D}_{\rm pre}, \mathcal{D}_{\rm post})$. 
\label{prop:bottleneck}
\end{prop}
\noindent In other words, as the number of histogram bins $L$ goes to infinity, the distributional difference $||\hat{p}_{{\rm pre}}-\hat{p}_{{\rm post}}||_2$ can be lower bounded by the bottleneck distance $d_B(\mathcal{D}_{\rm pre}, \mathcal{D}_{\rm post})$. This, combined with Equation (7) of the main paper, suggests that the greater the topological difference is between pre- and post-change data, the smaller its detection delay, which is as desired.

\end{appendices} 

\end{document}